\documentclass[preprint]{aastex}

\usepackage{graphicx}
\usepackage{natbib}
\def\icarus{Icarus}

\def\degree{\ifmmode {^\circ}\else {$^\circ$}\fi}

\newcommand{\lae}{\lower 2pt \hbox{$\, \buildrel {\scriptstyle <}\over {\scriptstyle\sim}\,$}}
\newcommand{\gae}{\lower 2pt \hbox{$\, \buildrel {\scriptstyle >}\over {\scriptstyle\sim}\,$}}
\def\sgn{{\rm sgn}}

\def\eqnum[#1]{(\ref{#1})}
\def\mysubsubsection#1{\subsubsection{#1}}

\def\Msolar{{\rm M}_\odot}
\def\Mearth{{\rm M}_\oplus}
\def\mearth{\ifmmode {\rm M_{\oplus}}\else $\rm M_{\oplus}$\fi}
\def\mjup{\ifmmode {\rm M_{J}}\else $\rm M_{J}$\fi}
\def\msun{\ifmmode {\rm M_{\odot}}\else $\rm M_{\odot}$\fi}

\def\powlaw{{\rm n}} 
\def\Psynodic{T_{\rm syn}}

\def\rhill{r_{\rm H}}
\def\rxing{r_{\rm xing}}

\def\da{\delta \tilde{a}}
\def\dadt{\frac{da}{dt}}
\def\dadtfast{\frac{da_{\rm fast}}{dt}}

\def\dadtembedded{\frac{da_{\rm emb}}{dt}}
\def\dadtgasI{\frac{da_{\rm I}}{dt}}

\def\dr{\delta r}
\def\dR{\delta R}

\def\ehill{e_{\rm H}}

\def\tmi{{\tau}}
\def\msafe{{m_{\rm safe}}}
\def\moli{{m_{\rm oli}}}
\def\miso{{m_{\rm iso}}}
\def\mfast{{m_{\rm fast}}}
\def\merode{{m_{\rm erode}}}
\def\a0a{\left(\frac{a_0}{a}\right)}

\begin{document}

\title{Migration of planets embedded in a circumstellar disk}

\author{Benjamin C. Bromley}
\affil{Department of Physics \& Astronomy, University of Utah, 
\\ 115 S 1400 E, Rm 201, Salt Lake City, UT 84112}
\email{bromley@physics.utah.edu}
\author{Scott J. Kenyon}
\affil{Smithsonian Astrophysical Observatory,
\\ 60 Garden St., Cambridge, MA 02138}
\email{skenyon@cfa.harvard.edu}

\begin{abstract}

Planetary migration poses a serious challenge to theories of planet
formation. In gaseous and planetesimal disks, migration can remove 
planets as quickly as they form.  To explore migration in a planetesimal 
disk,  we combine analytic and numerical approaches.  After deriving
general analytic migration rates for isolated planets, we use $N$-body 
simulations to confirm these results for fast and slow migration modes.
Migration rates scale as $m^{-1}$ (for massive planets) and 
$(1 + (\ehill/3)^3)^{-1}$, where $m$ is the mass of a planet and 
$\ehill$ is the eccentricity of the
background planetesimals in Hill units.  When multiple planets stir 
the disk, our simulations yield the new result that large-scale migration 
ceases. Thus, growing planets do not migrate through planetesimal disks.
To extend these results to migration in gaseous disks, we compare physical 
interactions and rates. Although migration through a gaseous disk is an 
important issue for the formation of gas giants, we conclude that 
migration has little impact on the formation of terrestrial planets.

\end{abstract}

\subjectheadings{planetary systems -- solar system: formation --
stars: formation -- circumstellar matter}

\maketitle

\section{Introduction}

Migration is an important physical process in planet formation 
\citep[e.g.,][and references therein]{lin86,war97,art04,lev07,pap07,kir09,dang2010b,lub10}.
Based on analytic theory and detailed numerical simulations,
several modes of interaction between growing planets and density
perturbations within a disk of gas or within a disk of planetesimals 
produce secular evolution of the orbital semimajor axis, eccentricity, 
and inclination of a planet.  For planets with masses exceeding
$\sim$ 0.1 \mearth, derived migration rates have a broad range, 
$\sim$ $10^{-7} - 10^{-4}$~AU~yr$^{-1}$. On typical timescales of 
0.1--1 Myr, planets can migrate through the entire disk.

To explain the frequency of ice giant and gas giant planets close to 
their parent stars, migration is essential \citep{lin96,mar02,ida04,ali04}.
Although there are significant selection biases, most known exoplanets
have semimajor axes, $a \lesssim$ 0.1--1 AU (data from exoplanet.org
and exoplanet.eu). Protostellar disks probably do not have enough mass 
to produce ice giants or gas giants so close to their parent stars 
\citep[e.g.,][]{bod00,kor02}.  Once these planets form farther out 
in the disk, however, they can slowly migrate inward to close-in orbits 
around their parent stars \citep[e.g.,][]{ida05,arm2007,tho08,mor09}.

Migration may also explain the orbital architecture of the Solar System.  
Observations of the dynamical structure of the Kuiper belt suggest that
Neptune migrated outward from its likely birthplace \citep{mal93,hah99}.
Other evidence suggests that the four gas giants formed in a more compact
configuration and then migrated outward \citep[e.g.,][]{tho02,tsi05,mor08}.

Despite these successes, migration is a great challenge for theories of 
planet formation. In the current picture, terrestrial planets and the cores 
of at least some gas giant planets form by a coagulation process, where 
lower mass objects collide and merge into larger objects. Early on, migration
timescales are long.  Without straying too far from their birthplaces,
protoplanets undergo runaway growth -- where a few of the largest objects 
grow much much faster than other objects -- and then oligarchic growth --
where these largest objects grow more slowly but still faster than much 
less massive objects \citep[e.g.,][and references therein]{kok98,gol04,kb10}.  
As planets begin to reach masses of $\sim$ 0.1~\mearth, however, collision 
times become longer than migration times. Thus, theory predicts that the 
final building blocks of planets migrate into the central star before they 
reach the mass of the Earth \citep{lin79,gt80,art93b,war97,mas03,ida08}. 

Migration is also a severe problem for the formation of ice giant and gas 
giant planets. Once ice giants or gas giants are fully-formed, migration 
can produce the close-in giant planets observed around nearby stars
\citep{ida05}.
However, theory predicts a more rapid migration of the lower mass building
blocks of ice and gas giants \citep{war97,mas03,ida08,pep08a}.  In the standard
theory, these lower mass planets migrate too fast to produce ice or gas 
giants.  Solving this problem is a central issue in planet formation theories.

Theories of migration generally focus on isolated planets interacting with 
the disk \citep[see][and references therein]{pap07}. Recent attempts to
understand how real planets avoid migration concentrate on the physics of 
this isolated interaction, including disk dynamics \citep[][2009b]{mas06,pep08a,paa09a}, 
magnetic fields \citep{ter03}, orbital eccentricity \citep{pap00}, 
disk thermodynamics 
(\citealt{kle08}, \\
\citealt{paa06b},
\citealt{paa08}, \citealt{kle09}, 
\citealt{paa10}, \citealt{paa11}), 
and 
turbulence \citep{nel04,ada09}. 
While any or all of these processes may reduce migration rates to acceptable 
levels, growing protoplanets are not isolated.  Tightly packed protoplanets 
probably perturb the disk differently than systems of widely spaced protoplanets. 
Thus, migration rates may depend as much on the local density of protoplanets 
as on the scale of specific interactions between an isolated planet and the disk.

Here, we consider how migration operates in systems of multiple planets.
Building on previous work \citep[e.g.,][]{mal93,hah99,lev07,kir09}, we 
examine migration in disks of planetesimals with single planets (\S2.1--2.3) 
and multiple planets (\S2.4--\S2.5). These results show that migration 
is rarely important in planet-forming disks of planetesimals. In \S3, we 
then explore the implications of our results for (inviscid) planetesimals 
embedded in (viscous) gaseous disks.  If our assumptions about viscous 
disks are valid, migration is rarely important during terrestrial planet 
formation. However, it is still an important issue in the formation of ice giant 
and gas giant planets.  We conclude with a brief summary and suggestions 
for further study in \S4.

\section{Planetary migration in a planetesimal disk}

Planets migrate through a planetesimal disk as a result of pairwise
exchange of angular momentum between the planet and individual disk
particles \citep{lin79,gt79,gt80,art93a}. An important distance scale for
this exchange is the planet's Hill radius,
\begin{equation}\label{eq:rhill}
\rhill = a \left(\frac{m}{3M}\right)^{1/3} , 
\end{equation}
where $a$ is the planet's semimajor axis, $m$ is its mass, and $M$ is
the mass of the central star. If the semimajor axis of a disk particle
is $a+\dr$, where $\dr$ is its orbital separation from the planet, then 
a passing encounter changes the planet's semimajor axis by
\begin{equation}\label{eq:scale}
\da \approx g(x) \frac{\rhill}{m} , 
\end{equation}
where $x = \dr/\rhill$, $g(x)$ is a function that depends on the
geometric shape of the planetesimal's trajectory relative to the
planet, and the tilde symbol indicates a change in orbital distance
per unit planetesimal mass.  Equation~(\ref{eq:scale}) asserts
that the dimensions of a particle's trajectory near the planet scale
as $\rhill$; the planet's recoil conserves momentum and must depend
on $1/m$.  

To calculate the trajectory function $g(x)$, we consider nearby particles 
in the co-orbital zone of the planet and more distant 
particles in the small-angle limit.  Planetesimals in the co-orbital 
zone, with $|\dr| \lae 2\rhill$, follow almost the same orbit as the 
planet but get pushed gently towards and away from it on horseshoe 
orbits \citep{der81}.  More distant planetesimals at $|\dr| \gae 4\rhill$ 
stream by the planet and experience small-angle scattering relative to 
their Keplerian path.  The trajectory function in these two cases is
\begin{equation}\label{eq:gx}
g(x) = \left\{
\begin{array}{cl}
      2x & \mbox{($|x| \lae 2$; \small co-orbital)} ,\\
 -32x^{-5} & \mbox{($|x| \gae 4$; \small small-angle scattering)} .
\end{array}
\right.
\end{equation}
The co-orbital zone result follows from conservation of energy. 
When the pair's relative speed is much greater than the planet's 
escape velocity at closest approach, the small-angle expression for 
larger separations follows from two-body scattering theory \citep{lin79}.

To illustrate this scaling property, we consider a set of numerical 
simulations of planetesimals on circular orbits close to a much more 
massive planet \citep[similar to Fig.~5 in][]{ida00}.  \citet{bk06}
describe our orbit integrator \citep[see also][]{bk10}. The planet
and the planetesimal start 180\degree\ out of phase on circular orbits
at distances $a$ (planet) and $a + \delta r$ (planetesimal) from the 
central 1 \msun\ central star.  We measure $da$ as the change in $a$ 
when the planet and the planetesimal complete a single synodic orbit. 

\begin{figure}[htb]
\centerline{\includegraphics[width=5.5in]{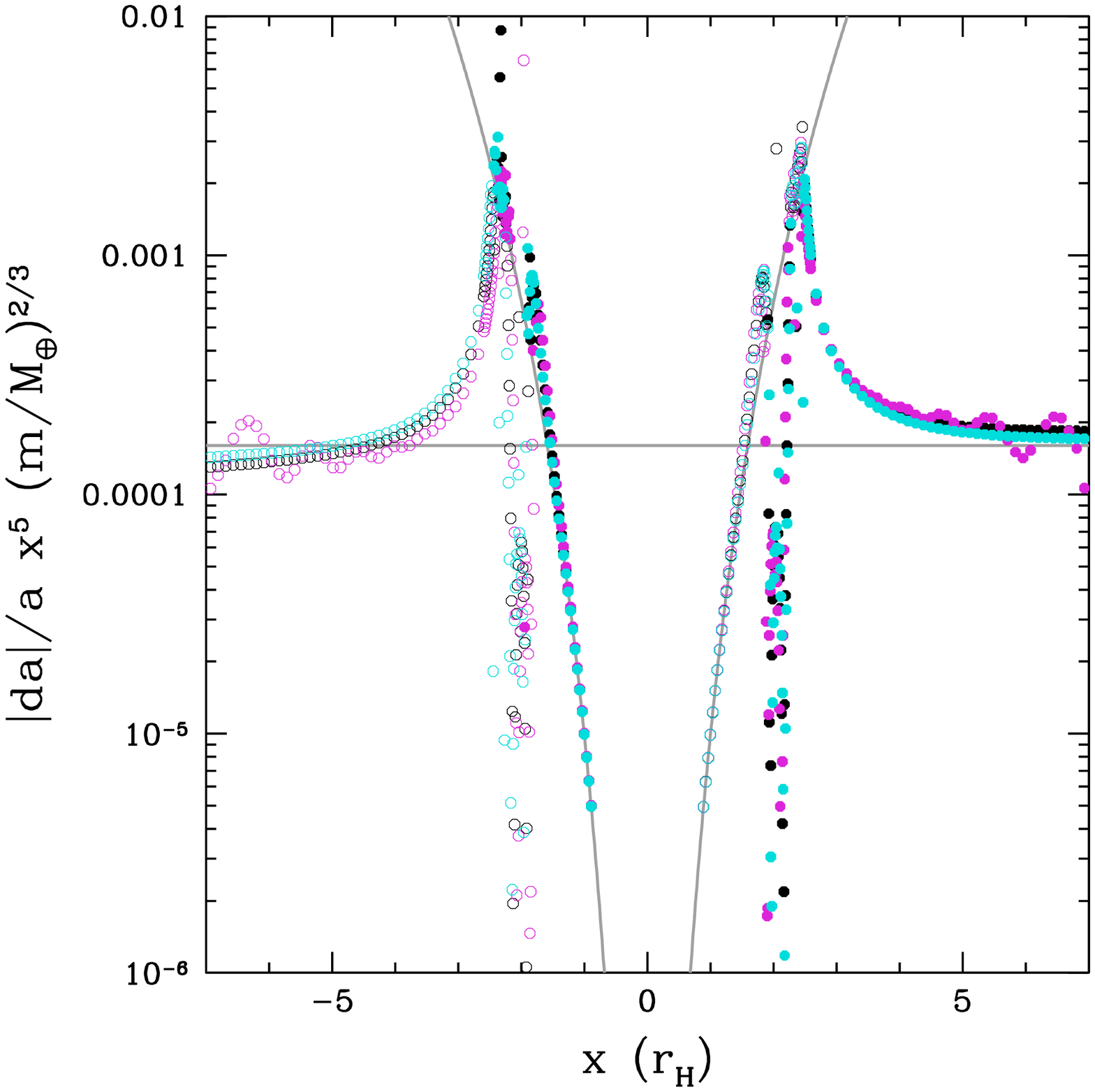}}
\caption{Derived change in semimajor axis ($da$) of a planet after 
  an encounter with a planetesimal as a function of their initial orbital 
  separation ($x$) in Hill units.  Objects start 180\degree\ out of phase on 
  circular orbits, with the planet at $a = 1$~AU from the central star 
  (1~\msun); $da$ (scaled by $x^5$ in the plot) is the resulting change 
  in $a$ after one synodic period.  Planetesimals have masses of 
  $5 \times 10^{-4}$ \mearth. Colors distinguish planet mass: 
  $m = 0.125$~\mearth\ (cyan), 1~\mearth\ (black), and 8~\mearth\ (magenta); 
  symbol attribute identifies the sense of migration: outward (open) or 
  inward (filled).  The scaling of $da$ with $m$ agrees with eq.~(\ref{eq:scale}). 
  The steep curves are theoretical predictions for the co-orbital zone 
  (eq.~[\ref{eq:gx}]); the horizontal line is from small-angle scattering theory.
\label{fig:da}}
\end{figure}

Our results agree very well with the scaling law (Fig.~\ref{fig:da}).
For three planet masses, scaled according to eq.~(\ref{eq:scale}),
the calculated $\da$ tracks the prediction well for co-orbital particles
($|x| \lae 1.8$) and distant particles in the small-angle limit ($|x| \gae 4$).
This scaling law begins to break down when the mass of the planet approaches 
the mass of the central star, but our results lend strong support for the 
``universality'' of the trajectory function $g(x)$. 

Orbital separations too distant for co-orbital encounters and too close 
for small-angle scattering encounters lead to chaotic scattering. A formal 
outer boundary for this limit is 
\begin{equation}\label{eq:rxing}
\rxing = 2\sqrt{3}\rhill .
\end{equation}
Outside this separation, there is an energy-angular momentum barrier that 
prevents chaotic orbit crossings for bodies on initially circular orbits 
\citep{gla93}. The inner boundary is the edge of the co-orbital region; thus, 
we adopt a chaotic zone with $1.8 ~ \rhill ~ \lae ~ \dr ~ \lae ~ 3.5 ~ \rhill$.
Fig.~\ref{fig:da} shows that particles in this region do not follow 
a simple trajectory function as in eq.~(\ref{eq:gx}).

\begin{figure}[htb]
\centerline{\includegraphics[width=5.5in]{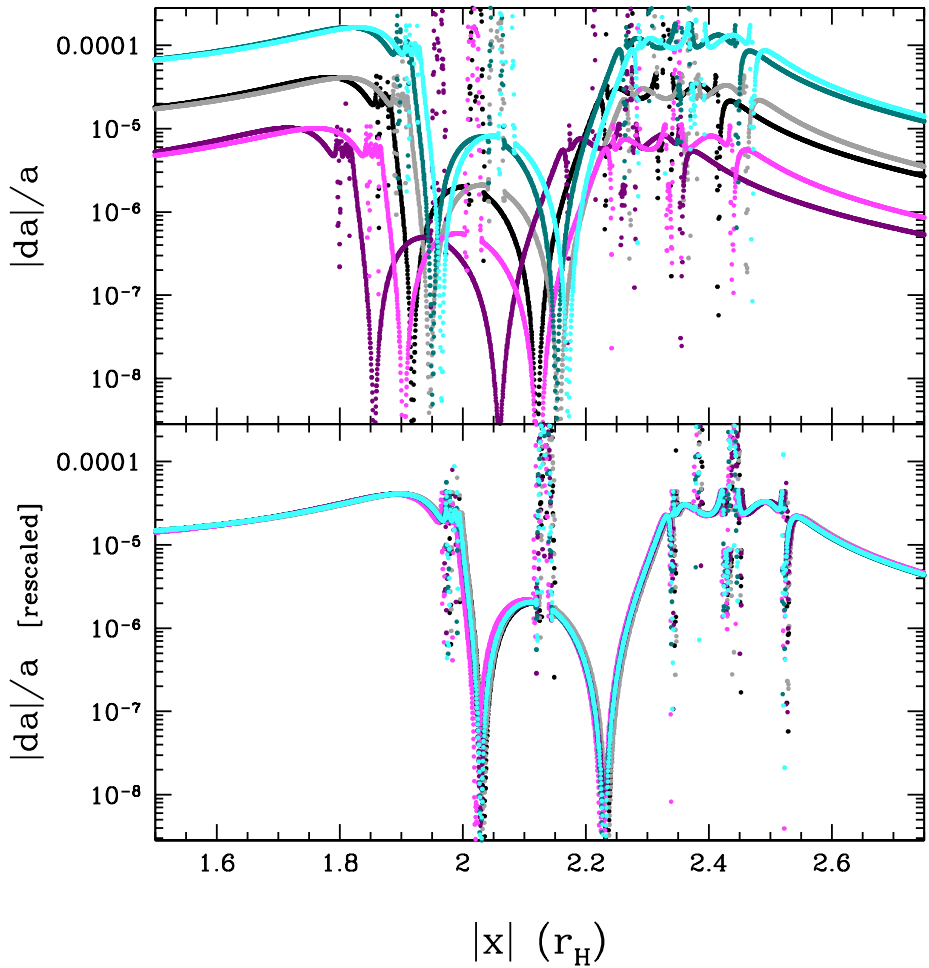}}
\caption{%
  As in Fig.~\ref{fig:da} for planetesimal orbits in the chaotic regime.  Colors 
  indicate mass for 0.125 (cyan), 1 (gray/black), and 5 (magenta) \mearth\ planets.
  The darker (lighter) shades indicate interactions with planetesimals that 
  are initially inside (outside) of the planet's orbit. The lower panel shows 
  the alignment of these curves after applying the scaling relation in eq. 
  (\ref{eq:shift}), which transforms $da$ into the ``universal'' trajectory 
  function, $g(x)\rhill/m$.
\label{fig:dachaos2}}
\end{figure}

The trajectory of a particle passing by a planet depends on whether the 
particle's approach is inside or outside the orbit of the planet.  This 
asymmetry is evident in Fig.~\ref{fig:da}, where particles with large 
negative $x$ have smaller $|da|$ than particles with large positive $x$.  
We can correct eq.~(\ref{eq:scale}) for this property of the orbits
using a first order Taylor series expansion:
\begin{equation}\label{eq:shift}
d\tilde{a} \approx  \left[g(x)-\frac{\beta x^2\rhill}{a} 
\frac{dg}{dx}\right]\frac{\rhill}{m} ~ .
\end{equation}
From numerical simulations, we estimate $\beta = 3/8$ in the chaotic regime
and $\beta = 9/20$ in the small-angle limit. Fig.~\ref{fig:dachaos2} 
shows $\da$ in simulations with various planet masses and with planetesimals 
that start inside and outside of the planet's orbit.  After normalizing our
results using eq.~(\ref{eq:shift}), these traces yield nearly the same 
``universal'' curve $g(x)$.

\subsection{Theoretical migration rates}

To estimate a migration rate from this formalism, we need a relation for
the encounter frequency. Although this frequency can vary substantially 
between consecutive passes of the same planetesimal \citep{kir09}, a good
characteristic number is the inverse of the synodic period,
\begin{equation}\label{eq:Psynodic}
\frac{1}{\Psynodic} \approx \frac{3|\dr|}{2 a T} 
\left(1 - \frac{5\dr}{4a}\right) \ ,
\end{equation}
where $T =2\pi a^{3/2}/(GM)^{1/2}$ is the orbital period of the
planet.  The product of this expression and eq.~(\ref{eq:shift})
yields a migration rate per unit mass of planetesimals. 

Extending this rate to a disk of planetesimals passing by the planet
requires a surface density distribution for the disk.  We adopt a smooth
surface density $\Sigma$ over an annulus with area $2 \pi r \dr$. 
Expanding all terms with $r = a + \dr$ in a Taylor series, keeping only 
first-order terms in $\dr$, and converting to a form with $x$ and $dx$
yields an integral for the migration rate:
\begin{eqnarray}\label{eq:dadt}
\nonumber
\dadt & = & \frac{\pi a^2 \Sigma}{M} \frac{a}{T} \int |x| g(x) \, dx \ \times
\\ & \ & \ 
 \left[1 + 
{
   \left(\frac{a}{\Sigma}\frac{d\Sigma}{da} - \beta\frac{x}{g}\frac{dg}{dx}
     - \frac{1}{4}\right) \frac{\rhill x}{a}
}
\right] .
\end{eqnarray}
where $g(x)$ is from eq. (\ref{eq:gx}). The surface density $\Sigma$ is 
often parameterized as a power-law, with
\begin{equation}\label{eq:Sigma_powerlaw}
\Sigma(a) = \Sigma_0 \left(\frac{a_0}{a}\right)^\powlaw ,
\end{equation}
and $\powlaw=1$--1.5. We use this form of $\Sigma$ throughout,
setting $\powlaw = 1$, $a_0 =1$~AU, and $\Sigma_0 = 30$~g~cm$^{-2}$
unless otherwise specified. 

\subsubsection{Migration from small-angle scattering}

To understand the implications of eq.~(\ref{eq:dadt}), we consider 
several simple cases.  For distant encounters between a planet and material 
in a power-law disk at separations $|x| ~ \gae ~ 4$, we derive the migration 
rate from eq.~(\ref{eq:dadt}) with $g(x) = -32/x^5$ and $\beta = 9/20$:
\begin{eqnarray}\label{eq:dadtffgam}
\nonumber
\frac{da}{dt} & = & -\frac{32 \pi a^2 \Sigma}{M} \frac{a}{T} 
   \int  \sgn(x) \frac{dx}{x^4} \ \times
\\ \mbox{} & \mbox{} & 
\ \ \ \left[1 + (2 - \powlaw)\frac{\rhill x}{a}\right] 
\ \ \ \ \ \ \ \ \ \ \ (|x|\gae 4),
\end{eqnarray}
where $\Sigma$ is evaluated at the planet's position.  If a planet lies 
embedded in a large disk where the inner (outer) radii are small (large) compared 
to the planet's semimajor axis, the first term in the square brackets vanishes;
the migration rate then depends weakly on planet mass through the second
term involving $r_H$ \citep[see also][]{ida00, kir09}. If a planet lies 
on the inside or the outside of the disk, the first term dominates. 

Providing it is no closer than $\rxing$ from the planet, eq.~(\ref{eq:dadtffgam}) 
predicts that a disk situated just inside or outside a planet's orbit is repulsive. 
To confirm this behavior numerically, we use an $N$-body code that evolves massive 
planets, along with massive planetesimals that interact with the planets but not 
with each other \citep[i.e., the disks are not self-gravitating; see][]{bk10}.  
For example, a 100~$\Mearth$, $\powlaw=1$ power-law disk consisting of $2\times 10^5$ 
equal-mass particles between 26.5~AU and 35.5~AU pushes a 0.3~$\Mearth$ planet at 
25~AU inward with a speed of 0.012~AU/10~kyr (eq.~\ref{eq:dadtffgam}). 
Simulations of 100 planetary orbits yields $0.0117\pm 0.004$~AU/10~kyr.
Tests with a disk inside the planet's orbit confirm that the planet migrates
outward, as expected from eq.~(\ref{eq:dadtffgam}).

With this formalism, we can consider an idealized example of the migration of a 
planet nestled between two equal-mass annuli of planetesimals. For a surface 
density $\Sigma \propto a^{-1}$ and spacing between the planet and each annulus 
of $\dr > \rxing$, eq.~(\ref{eq:dadtffgam}) predicts a net inward migration.  
For an Earth-mass planet at 25~AU, between two 0.5~AU annuli centered on 23~AU 
and 27~AU, with 50~$\Mearth$ apiece, the theoretical migration rate is 
$-0.019$~AU/10~kyr.  Although the planet eventually migrates through the gap into 
the inner disk of planetesimals, our numerical simulation using 1/600~\mearth\ planetesimals 
yields a migration rate of $-0.025\pm 0.002$~AU/10~kyr. In this simulation, migration 
leads to more interactions with the inner disk of planetesimals than predicted by 
the analytic theory; still, this numerical result agrees reasonably well with the 
analytic prediction.

In these examples, the gap between the planet and the disk spans the chaotic and 
co-orbital zones.  In small-angle scattering, migration is fairly small, $\sim$ 
1--2~AU~Myr$^{-1}$. For typical growth times of $\sim$ 1--3 Myr
\citep[e.g.,][]{kb06,bk10}, planets migrate through a small fraction of the disk.

\mysubsubsection{Fast migration}

For a planet embedded in a planetesimal disk, the co-orbital zone is much 
more important than the small-angle scattering regime \citep{war91,ida00}.  
Over a complete libration period of a horseshoe orbit, there is no net migration 
of a planet responding to a planetesimal.  If the planet is already moving
radially inward (or outward) on a timescale shorter than the libration period, 
the situation is different.  The planet can then pull itself along, continually 
exchanging places with the co-orbital material in its path.  When this 
mechanism works, it is efficient and relatively fast.

Simulations performed with our code and other codes \citep{ida00,kir09} 
suggest that fast migration can be inward or outward.  \citet{kir09} identify 
a strong preference for inward migration. Our calculations confirm this
conclusion; more massive planets also seem to migrate inward more often than
less massive planets.  We speculate that inward migration dominates in 
most simulations from the gentle inward push of the weakly scattered disk, 
whose influence on a planet increases with $m$.

The fast migration rate $da_{\rm fast}/dt$ follows from integrating 
eq.~(\ref{eq:dadt}) over the half of the co-orbital zone that a planet traverses.
We adopt this half-width as $\delta r = X_{co} \rhill$, with $X_{co} = 1.8$. 
However, fast migration 
occurs only if the rate allows a planet to clear the co-orbital zone during 
the libration period of the planetesimal at the zone's edge. Otherwise,
the planetesimal orbits back and provides a counter-torque before the planet
migrates away.  Large planetary 
masses have large co-orbital zones that are hard to traverse in a single
libration period. Thus, this requirement sets a mass limit on fast migration,
\begin{equation}\label{eq:mfast}
\mfast \approx 4.0 ~ \left(\frac{2 \pi a^2 \Sigma}{3M}
\frac{X_{\rm co}}{1.8}
\right)^{3/2} M . 
\end{equation}
In a disk with a surface density of 30~g~cm$^{-2}$ at 1 AU from a solar mass 
central star, this limit is $\mfast \approx 0.025$~\mearth. 

To estimate the migration rate for planets more massive than $\mfast$, we consider 
a simple model.  Fast migration relies on a planet crossing the co-orbital zone,
with size $\dr \sim \rhill$, within a typical synodic period of an orbiting planetesimal. 
When $m > \mfast$, the co-orbital zone is too large for the planet to cross in a 
single synodic period. Thus, a fraction of the material in the co-orbital zone has 
multiple interactions with the planet, slowing the migration rate. This fraction 
increases with $m$; thus, more massive planets migrate more slowly. To quantify this 
statement, we define $X_{co,fast}$ as the size of the co-orbital zone for a planet 
with $m = \mfast$. Planets with $m > \mfast$ have larger co-orbital zones, with 
$X_{co} > X_{co,fast}$.  For these planets, we assume that planetesimals within 
a distance $X_{co,fast}$ of the planet contribute to migration; co-orbiting 
planetesimals beyond this distance do not contribute.  Integrating over this annulus, 
as in eq.~(\ref{eq:dadt}), and using $\rhill \propto m^{1/3}$, the attenuation 
factor scales as $\mfast/m$. For $m > \mfast$, migration rates scale inversely with 
the mass of the planet\footnote{Using a different approach, \citet{war91} notes that
migration saturates when the planet cannot drift across the co-orbital zone in a synodic 
period \citep[see also][]{paa09a}. Our derivation yields the mass dependence directly.}.

Fig.~\ref{fig:migratemass} illustrates several numerical simulations of fast 
migration for planets with a broad range of masses. Following \citet{kir09},  
each planet lies embedded in a power-law disk extending from 14.5~AU to 35.5~AU
with $\Sigma = 1.2~(a /{\rm 25~AU})^{-1}$ g~cm$^{-2}$.  We represent the disk with 
particles that are each 1/600$^{\rm th}$ of the mass of the planet.  To speed up the 
onset of fast migration, the co-orbital zone ($\delta r \le \rhill$) is initially
clear of particles.
We also scale the r.m.s.\ planetesimal eccentricity to keep the same initial 
$e = \rhill/a$ for each planet. The nearly identical migration tracks for masses 
below $\mfast \approx 3~\mearth$ illustrate fast migration at the theoretical 
rate indicated by the dashed curve.  When the planet encounters the inner edge 
of the disk, the rates fall to zero (and sometimes reverse sign).  More massive 
planets follow tracks that reflect the $1/m$ attenuation of the fast migration 
rate for $m > \mfast$. 

Several aspects of protoplanetary disks conspire to set limits on the {\it minimum} 
planet mass for fast migration. In a gaseous disk, small planetesimals with 
$St < \alpha$ are entrained in the gas, where $St = r \rho_g \Omega / \rho c_s$ is
the Stokes number, $r$ and $\rho$ are the radius and mass density of a planetesimal, 
$\rho_g$ is the local gas density, and $c_s$ is the sound speed 
\citep[see][and references therein]{you2007,chi2010,orm2010}.  Fast migration 
requires that the Hill radius of the planet exceed the scale height, $h_s$, of the 
planetesimals.  Following \citet{you2007}, $h_s = h ~ {\rm min}(1, \sqrt{\alpha/St})$, 
where $h$ is the scale height of the gas and $\alpha$ is the disk viscosity parameter. 
Adopting a simple expression for the disk scale height, $h = h_0 (a / {\rm 1~AU})^{9/7}$ 
\citep[e.g.,][]{kh87,chiang1997} and requiring $h_s < \rhill$ yields a simple expression 
for the minimum mass for fast migration in a gaseous protoplanetary disk:
\begin{equation}
\label{eq:mfast-min-gas}
m_{fast,min} \gtrsim 36 f_{st} \left ( \frac{h_0}{0.033} \right )^3 \left ( \frac{a}{\rm 1~AU} \right )^{3/4} ~ \mearth\ ,
\end{equation}
where $f_{st} = {\rm min}(1, (\alpha/St)^{3/2})$. When $f_{st}$ = 1, eq.~(\ref{eq:mfast-min-gas})
yields an approximate condition for fast (type III) migration through the gaseous disk 
\citep[e.g., eq. \ref{eq:gap}; see also ][2008b] {mas03, dang2005, cri2006, pep08a}.  
When most of the solid material is in much larger particles with $St \gg 1$, lower mass 
planets undergo fast migration through the planetesimals.  For 1 km planetesimals with 
$St \sim 10^3$ and $\alpha = 10^{-2}$, $m_{fast,min} \approx 10^{-6}$ \mearth\ at 1 AU. 

In a planetesimal disk, particle growth sets another limit on the minimum mass for fast 
migration. During oligarchic growth, leftover planetesimals have typical velocity dispersions,
$v \approx \epsilon v_{esc}$, where $v_{esc}$ is the escape velocity of the largest oligarch
and $\epsilon \approx \Sigma_o / \Sigma_s$ is the ratio of the surface density in oligarchs to 
the surface density in planetesimals \citep[e.g.,][2010]{gol04,kb08}. The scale height of the
planetesimals is then $h_s = v \Omega^{-1} \approx \epsilon v_{esc} \Omega^{-1}$, leading
to a simple expression for the ratio of the scale height to the Hill radius in a disk
surrounding a solar-type star:
\begin{equation}
\label{eq:scale-height}
\frac{h_s}{\rhill} \approx 20 \rho^{1/6} \epsilon ~ .
\end{equation}
Thus, planets undergo fast migration through planetesimals only when they contain no more 
than $\sim$ 5\% ($\epsilon \lesssim$ 0.05) of the mass in solid material.

\mysubsubsection{Migration rate summary.}

Here, we summarize the migration rates calculated from eq.~(\ref{eq:dadt}) 
for fast migration (with the reduction factor for large masses), and for a planet 
embedded in a disk that moves relatively slowly through small-angle scattering:
\begin{eqnarray}
\label{eq:dadtfast}
\dadtfast & = & 
\medskip
\displaystyle
\pm 3.9 ~ \frac{\pi a^2 \Sigma}{M} 
\left(\frac{X_{co}}{1.8}\right)^3
{\rm min}\left(1,{\mfast}/{m}\right) \, \frac{a}{T}\ ,
\\
\label{eq:dadtembedded}
\dadtembedded & = & 
\bigskip
\displaystyle
-32 (2-\powlaw) ~ \frac{\pi a^2 \Sigma m}{3 M^2}~\frac{a^2}{\dR^2}\,\frac{a}{T}\ ,
\medskip
\\
\label{eq:dadtembedded2}
\mbox{} & = & -\frac{8}{3}(2-\powlaw) ~ \frac{\pi a^2 \Sigma}{M}~\left ( \frac{m}{3M} \right )^{1/3} \frac{\rxing^2}{\dR^2}\,\frac{a}{T}\ ,
\end{eqnarray}
where $\dR$ is the physical distance separating the planet and the
edge(s) of the disk. The $\pm$ sign for fast migration indicates it
can be either inward or outward, at least for small mass planets.
In each case, we have kept only leading order terms in $\rhill/a$ and 
have assumed that the inner and outer edges of the disk are far, far 
away from the planet.

In a thorough analysis of the orbits of moonlets embedded in Saturn's ring 
system, \citet{cri2010} derive $g(x)$ in the chaotic regime. Their eq. (39) 
for the migration rate has the same functional form as our eq.~(\ref{eq:dadtembedded2}), 
including the $m^{1/3}$ dependence. Because we 
derive rates in the small-scattering limit, the numerical coefficient in 
eq.~(\ref{eq:dadtembedded2}) is a few times smaller than the equivalent 
coefficient in the \citet{cri2010} rate. Considering the differences in the
two approaches, the agreement in the functional form and the magnitude of 
the migration rate is encouraging.


The migration timescale is $\tmi \equiv a/|da/dt|$. For a planet embedded in
a power-law disk (eq.~[\ref{eq:Sigma_powerlaw}]), the fast migration mode and 
slow, embedded migration yield
\begin{eqnarray}\label{eq:taufast}
\tau_{\rm fast}({\rm yr}) & = & 2.3 \times 10^4 \ 
      \mbox{max}\left(1,{m}/{\mfast}\right) 
\\
\label{eq:tauemb}
\tau_{\rm emb}({\rm yr}) & = & 3.6 \times 10^6  \ 
       \frac{(\dR/a)^2}{0.035^2} \ 
       \left(\frac{m}{\mearth}\right)^{-1} 
   \\
\label{eq:tauemb2}
\displaystyle\medskip
       \mbox{} & = & 3.6 \times 10^6 \ 
       \frac{\dR^2}{\rxing^2} 
       \left(\frac{m}{\mearth}\right)^{-1/3} 
\end{eqnarray}
for fiducial parameters of $M = 1$~\msun, $a = 1$~AU, and $\Sigma_0 =
30$~g~cm$^{-2}$.  For other situations, these timescales vary as
\begin{equation}
\tau \propto a^{\powlaw-1/2} \Sigma_0^{-1} M^b
\label{eq:taupow}
\end{equation}
where $b = $ 1/2 for fast migration with $m<\mfast$, $b = $ 3/2 for
embedded migration at fixed $\dR/a$ (eq.~[\ref{eq:tauemb}]), and $b = $ 
5/6 for embedded migration at fixed $dR/\rxing$. The dependence on these
parameters is more complicated for attenuated fast migration, with $m
> \mfast$ (cf.~eq.~[\ref{eq:mfast}]).

Although fast migration is two orders of magnitude faster than the 
slow mode, embedded migration may sometimes dominate. In eqs.
(\ref{eq:tauemb}) and (\ref{eq:tauemb2}), $\dR$ is the distance between 
the planet and the nearby edges of the disk. By construction, $\dR > \rxing$. 
If the disk is dynamically warm, all interactions except for the small-angle 
scatterings are washed out. In a dynamically cold disk, a planet might make
its own gap by scattering away all but the more distant material 
\citep[e.g.,][]{raf2001}.  These two situations are similar to Type I and 
Type II migration in a gaseous disk (\S\ref{ssect:gaseousdisk}).

For many of these migration modes, the power law variation of $\tau$ with $a$
in eq.~(\ref{eq:taupow}) yields an integrable expression for $a(t)$. 
In all modes of fast migration and embedded migration with a constant or 
slowly varying ratio $\dR/a$ \citep[e.g.,][]{ale09},  we can adopt 
$da/dt = C a^{\gamma_1}$, where $\gamma_1 = 3/2 - \powlaw$, and derive a
simple expression for the time variation of the semimajor axis,
\begin{equation}\label{eq:a(t)}
a(t) = \left[
        -\frac{C t}{2^{\gamma_{2}-1}} + a(0)^{1/\gamma_2}
       \right]^{\gamma_{2}}
\end{equation}
where $\gamma_2 = 2/(2\powlaw-1)$. Standard models for the minimum mass
solar nebula \citep[e.g.,][]{wei77,hay81} have $\powlaw$ = 3/2; the semimajor
axis of the orbit then contracts linearly with time.  Radio observations of 
young stars are more consistent with $\powlaw$ = 1 \citep[e.g.,][]{and07,ise09}; 
$a$ then contracts quadratically with time. Fig.~\ref{fig:migrateecc} compares
our numerical simulations with the analytic results for $a(t)$.

To apply these results to more realistic disks with planets and planetesimals,
we consider several examples derived from our planet formation simulations. 
We begin in \S2.2 with disks stirred by growing protoplanets, continue in \S2.3
with disks declining rapidly in mass, and conclude in \S2.4 with disks containing
many growing planets.

\subsection{Stirred disks}

In the planetesimal theory, planets form by accreting smaller objects along
their paths. When planetesimals are large ($r \gtrsim 0.1$ km), growing planets 
gravitationally stir up the orbits of 
neighboring planetesimals \citep{art97,kl98,kb02}. Thus, growing planets
dynamically heat up the disk. In Hill units, with $\ehill \equiv e a/\rhill$, 
ensembles of growing planets rarely produce planetesimals with 
$\ehill ~ \lae ~ 5$, where $e$ is the orbital eccentricity of a planetesimal.  
During runaway growth, planets grow to masses of roughly 
0.001--0.01 \mearth; the r.m.s.\ eccentricities of planetesimals usually drop 
from $\ehill \sim 100$ to $\ehill \sim 5$.  Throughout oligarchic growth, 
$e \propto \epsilon v_{esc} / v_K$, where $v_{esc}$ is the escape velocity of the 
most massive planet, $v_K$ is the local circular velocity, and $\epsilon$ is the
ratio of the mass in oligarchs to the mass in planetesimals \citep[e.g.,][]{kl98,gol04,kb08}.
In Hill units, $\ehill \propto \epsilon v_{esc} / \rhill \propto \epsilon$. Thus, 
$\ehill$ grows slowly as oligarchs accrete more and more planetesimals, reaching 
$\ehill \sim 100$ during the late stages of oligarchic growth and throughout 
chaotic growth \citep[][2006, 2008, 2010]{kb04}.  Inclinations are typically 
half these values\footnote{In a disk where oligarchs grow from collisions with 
small planetesimals coupled to the gas, $e$ and $i$ are set through interactions
with the gas instead of stirring by oligarchs (\S2.1.2). When the gas produces
$e > \ehill$, migration slows as in eq. \ref{eq:migrateecc}.}.

In a hot disk, interactions between a planet and surrounding planetesimals weak, 
slowing the migration rate by a factor of 
\citep[see][]{ida00, kir09}
\begin{equation}\label{eq:migrateecc}
\dadt \gae \left. \dadt\right|_{\ehill=0} \left[1 + (\ehill/3)^{3}\right]^{-1}
\end{equation}
To test this prediction, we repeat the calculations for Fig.~\ref{fig:migratemass} 
and vary the r.m.s.\ value of the initial $\ehill$ for planetesimals from unity to 
50.  For larger values of initial $\ehill$, migration is undetectable in a $10^5$~yr 
time frame. For $\ehill \approx$ 10--50, it is a challenge to prohibit particles from
the co-orbital zone at the start of the simulation. To keep all calculations in this 
suite on the same footing, we allow planetesimals for all initial $\ehill$ to reside
in the co-orbital zone. The fraction of particles in the co-orbital zone is small; 
most are not on horseshoe orbits. Still, the onset of the fast migration mode is 
somewhat slow compared to the results in Fig.~\ref{fig:migratemass}.

The results of these simulations (Fig. \ref{fig:migrateecc}) follow the trend expected 
from eq.~(\ref{eq:migrateecc}).
Planets embedded in a disk of low eccentricity planetesimals migrate rapidly, at a rate
that scales inversely with the mass of the planet (see also Fig. \ref{fig:migratemass}).
As we raise the initial $\ehill$, the migration rate slows. For $\ehill \approx$ 50, the 
migration rate is negligible. As shown by the dashed lines in the upper panel of the figure,
the reduction in the migration scales roughly as $\ehill^{-3}$.

\subsection{Eroded disks}

As an individual planet migrates through a disk annulus, it disrupts the disk 
\citep{ida00,kir09}. In fast migration, a planet tosses material into relatively 
eccentric orbits in random directions.  This scattering process reduces the local
surface density of planetesimals.  If the planet moves fast enough to encounter 
only the unperturbed disk upstream from it, this disturbance has little adverse 
impact on the planet's migration rate.

For more massive planets, slower migration may lead to a continuous loss of disk 
material. Planets migrate more slowly in less massive disks.  To quantify this 
effect, we let the disk surface density within $\rhill$ of the planet vary as
\begin{equation}
\dot{\Sigma}(a,t) = -\frac{\epsilon}{\Psynodic} \Sigma(a,0)
\label{eq:sigmat}
\end{equation}
where $t$ is the time since a planet reaches a narrow annulus of the
disk at orbital distance $a$ and $\epsilon$ describes the efficiency
of a planet in scattering disk material. To derive the migration rate for
a disk with this exponentially decaying surface density, we integrate eq. 
(\ref{eq:sigmat}) over the time it takes the planet to migrate a distance $\rhill$
from eq.~(\ref{eq:dadt}).  Thus, we approximate the instantaneous surface 
density by a temporal and spatial average inside an active zone of width 
$O(\rhill)$, where $\epsilon \sim \rhill/a$ is an efficiency factor for 
clearing the zone of planetesimals.  This approximation leads to a non-linear 
equation where the average value of $\Sigma$ depends on the migration rate. 
Solving this equation leads to a new migration rate:
\begin{equation}\label{eq:diskerode}
\dadt \approx \frac{3\epsilon\rhill^2}{2a T} \left[
\ln\left(1-\frac{3\epsilon\rhill^2}{2 a \dot{a}_0 T}\right) 
\right]^{-1} ~ ,
\end{equation}
where $\dot{a}_0$ is the theoretical migration rate of the planet without any 
time-variation in the disk surface density.  The sensitivity of this model to 
the exact form of $\epsilon$ is fairly weak.

When the argument of the logarithm in eq.~(\ref{eq:diskerode}) is zero, 
the migration rate vanishes. Thus, this expression implies a high mass cut-off, 
$\merode$, where more massive planets cannot migrate through the disk.
From eq.~(\ref{eq:dadtfast}), this limit is
\begin{eqnarray}\label{eq:merode}
\merode & \sim & \sqrt{8\pi a^2\Sigma\mfast} \\ 
            & \sim & 0.8 \left(\frac{a}{1~\mbox{AU}}\right)^{7/4-\powlaw/2}
\ \mearth .
\end{eqnarray}
If disk erosion is an important process, it begins when a planet reaches 
roughly an Earth mass at 1~AU and over 40~\mearth\ at 25~AU. Removal of disk 
material by scattering is at least enhanced, if not entirely enabled, by 
secondary scattering from nearby planets. Migration in principle, can be 
virtually halted for higher mass planets if they have neighbors that prevent 
the return of scattered planetesimals.

\subsection{Multiple planets in a disk}

In the coagulation paradigm, planets grow hierarchically from smaller
planetesimals. As they grow, planets almost always have neighbors of
comparable mass. During runaway and oligarchic growth, a few large objects 
try to accrete all of the mass in an annulus. Once these oligarchs contain
roughly half of the total mass, they begin to interact chaotically 
\citep{gol04,kb06}. During chaotic growth, planets scatter planetesimals 
to large $\ehill$ and grow by large collisions with other planets \citep{kb06}. 
Once chaotic growth begins, smooth migration through a sea of planetesimals 
is impossible.  Thus, we consider migration in a disk of growing oligarchs 
which contain less than half of the mass in solid material.

Planets affect the migration of a neighboring planet in two ways. As a planet
migrates through a disk, it stirs up the planetesimals along its orbit.  After
migrating past these excited planetesimals, the planet leaves behind a wake of
planetesimals with large $\ehill$ \citep[see also][]{edg04,kir09}. This wake 
is a barrier that prevents other planets from migrating inward from larger $a$.  
For a planet migrating through a disk with initial $\ehill \approx$ 1, 
planetesimals left behind have typical $\ehill ~ \gae ~ 3 - 5$. From eq. 
(\ref{eq:migrateecc}), planets encountering stirred up planetesimals have factor 
of 2--6 times smaller migration rates. In addition, planets migrating into a 
wake require longer periods to clear their co-orbital zones of dynamically ``hot''
planetesimals. As a result of these factors, migration ceases.

Fig.~\ref{fig:pairfast} illustrates this phenomenon.  Two planets migrate 
inward in the fast migration mode; the migration of the outer planet stops
abruptly when it encounters the wake of planetesimals already stirred up 
by its partner.

Migrating planets can also deflect planetesimals that chaotically scatter 
from a neighboring planet. When a planet deflects planetesimals from its 
neighbor, it prevents the planetesimals from returning to the neighbor. The loss 
of these encounters prevents the neighbor from migrating towards the planet.
Thus, the two planets recoil from the material that is passed between them
\citep[see also][]{mal93,hah99}.  

To demonstrate this process, we simulate the migration of a Saturn-mass planet 
at 10~AU embedded in a massive disk \citep[400~\mearth\ between 6~AU and 20~AU, 
with a power law surface density distribution and $\powlaw = 1$; see][]{lev07}. 
We then vary the mass of a second planet at 5~AU.  As the mass of the inner 
planet falls from a jovian mass to 30~\mearth, the sense of migration of the 
outer planet changes from outward (Fig. \ref{fig:migrateJS}; black curves) 
to inward (blue curve).  Until it encounters scattered planetesimals from the
outer planet, the inner planet slowly migrates inward at the ``adjacent'' rate 
from eq.~(\ref{eq:dadtffgam}).  Despite its small Hill radius 
($\rhill = $ 0.155~AU), the 30~\mearth\ inner planet has a considerable impact 
on the migration rate of a much more massive outer planet 5~AU away. Migration 
is remarkably fragile.

To conclude this section, we consider migration in a multi-planet system. In
our simulations of planet formation \citep[e.g.,][]{kb06,bk10}, planets with masses 
of 0.1--1 \mearth\ are often separated by 10--20 mutual Hill radii. To investigate 
migration in an idealized version of these calculations, we simulate the 
evolution of six 0.5~\mearth\ planets in a disk of planetesimals extending 
from 7--35.5 AU. As in the calculations for Fig.~\ref{fig:migratemass}, 
the planetesimals have an initial surface density distribution $\Sigma = 
1.2 ~ (a / {\rm 25~AU})^{-1}$ g~cm$^{-2}$. 
Unlike the calculations in Fig.~\ref{fig:migratemass}, co-orbital zones are 
initially filled with planetesimals.

Fig.~\ref{fig:multi} summarizes the main results of these simulations. In a
multi-planet system, long-term migration rates are small. Initially, each planet
clears its co-orbital zone of material in $3 - 6 \times 10^4$ yr. Fast migration
commences. Eventually, each planet encounters the ensemble of stirred up planetesimals 
left behind by its inward neighbor. Migration stops.  In these examples, migration of 
the innermost planet ceases when it reaches the inner edge of the disk. In disks with
small inner radii, migration of the outer planets ceases well before the inner planet
reaches the inner edge of the disk.

\subsection{Migration and planet formation}

To place these results in the context of formation scenarios, we consider the growth 
and migration of planets in the planetesimal theory. In this picture, planetesimals 
ranging in size from $\sim$ 0.1~km to $\sim$ 100~km condense out of the gaseous disk.
Planetesimals collide and merge into larger and larger objects.  After short periods 
of orderly and runaway growth, the largest objects enter oligarchic growth, where
they continue to accrete and to stir up leftover planetesimals. During this phase, 
dynamical friction between planetesimals and oligarchs dominates dynamical interactions 
among oligarchs. Thus, oligarchs remain fairly isolated from one another.  Once 
oligarchs contain roughly half of the mass in solid material, their mutual dynamical
interactions dominate dynamical friction with planetesimals. Oligarchy ends. Chaotic 
growth, where oligarchs grow by giant impacts and continued accretion of small 
planetesimals, begins \citep{gol04,kb06}

During the transition from oligarchic to chaotic growth, the `isolation mass' sets 
the mass of the largest oligarchs \citep{lis87,kok98,gol04}. By definition, isolated 
objects have small dynamical interactions; thus, their typical separations are
$\sim B r_H$ with $B \approx$ 7--10 \citep[][2000, 2002]{lis87,kok98}.  When an object 
contains all of the mass in an annulus of width $B r_H$, it reaches the isolation mass. 
With $\miso = 2 \pi a \Sigma B \rhill$ and $\Sigma = \Sigma_0 a^{-n}$, 
$\miso = (2 \pi B \Sigma_0)^{3/2} ~ (3 M)^{-1/2} ~ a^{3 - 3n/2}$.  If we adopt a 
disk with $n$ = 1, $\Sigma_0$ = 10~g~cm$^{-2}$, and $B$ = 7, isolated objects have
separations of $2 \rxing = B \rhill$ and lie well outside the co-orbital zones of 
their nearest neighbors.  The isolation mass is then\footnote{Our definition is 
appropriate for the onset of chaotic growth, when $\ehill > 1$; when $\ehill < 1$, the 
alternative of \citet{gol04} provides a better measure of the masses of isolated objects.}
\begin{equation}
\label{eq:miso}
\miso = 0.07 \left ( \frac{\Sigma_0}{\rm 10~g~cm^{-2}} \right )^{3/2} ~ \left ( \frac{a}{\rm 1~AU} \right )^{3/2} ~ \left ( \frac{1~\msun}{M} \right )^{1/2} ~ \mearth ~ .
\label{eq:m-iso}
\end{equation}

For each oligarch, the ratio of $\miso$ to $\mfast$ (eq.~[\ref{eq:mfast}]) sets
the importance, the mode, and the timing of migration through a sea of planetesimals.  
Low mass oligarchs with $\moli < \miso$ can migrate, but they cannot migrate freely.  
The typical radial spacing of low mass oligarchs is 
$r_{oli} \approx 7 (\moli / \miso)^{2/3} \rhill$.  With 
$r_{oli} \ll 7 \rhill$, an oligarch migrates only a few $\rhill$ before it
encounters the wakes of other oligarchs. Migration then ceases.  Once massive oligarchs 
have $m \gtrsim \miso$, they are free to migrate. However, massive oligarchs also interact
chaotically. At 1--10 AU, the timescale for chaotic growth is shorter (longer) than the 
timescale for slow (fast) migration. When $\miso < \mfast$, migration is important
during chaotic growth. Otherwise, growing oligarchs do not migrate through a sea
of planetesimals.

This analysis suggests that migration through an ensemble of planetesimals is rarely 
important within planet-forming disks.  From our definitions of $\miso$ and $\mfast$ 
(eq.~[\ref{eq:mfast}]), the ratio $\miso / \mfast = 3 B^2 / 4$ $\approx$ 37. 
Although growing oligarchs migrate in the fast mode, they can never migrate very
far before they encounter the wake of another migrating oligarch. Migration then
ceases (Fig.~\ref{fig:pairfast} and Fig.~\ref{fig:multi}). 

There are several plausible exceptions to this conclusion.  If collisional damping 
or gas drag reduce the $e$ and $i$ of stirred up planetesimals in the wake of a 
migrating planet, then another planet can migrate through the wake.  When the wake 
consists of large planetesimals with $r \gtrsim$ 0.1~km, however, collisional damping 
and gas drag are ineffective. Dynamical friction and viscous stirring by large 
planetesimals and small oligarchs keep particles at large $e$ and $i$ 
\citep[e.g.,][]{kl98,gol04}.  Thus, oligarchs cannot migrate freely through a disk 
of large planetesimals. For smaller particles, damping may reduce $e$ and $i$ on 
timescales comparable to the migration timescale \citep[e.g.,][]{kb01}.  Because 
damping and migration occur on similar timescales, closely-spaced oligarchs probably 
encounter wakes before damping can smooth them out. Widely-spaced oligarchs suffer 
chaotic growth, which keeps planetesimals stirred up despite damping. Thus, we 
conclude that damping does not allow migration in a planetesimal disk.

Scattering may also lead to effective migration \citep[][2010]{lev07,ray09a}.  
During chaotic growth, 
massive planets scatter lower mass planets farther out in the disk 
\citep[e.g.,][]{mar02,ver09,ray10,bk10,cha10}. 
At large $a$, oligarchs form slowly \citep[e.g.,][2010]{kb08}. Thus, planets formed 
at small $a$ and scattered to large $a$ may end up in a calm disk composed of
planetesimals with small $\ehill$.  Without other oligarchs to impede them, these 
scattered planets can then migrate freely through the outer disk. 

Once chaotic growth ends, any leftover planetesimals can support inward or outward 
migration. For leftovers with large $e$ and $i$, migration rates are slow. However,
outwardly migrating planets may reach planetesimals with much lower $e$ and $i$,
enhancing migration rates.  Several dynamical models for the origin of the Solar 
System rely on migration through a leftover planetesimal disk 
\citep[e.g.,][]{hah99,tsi05}. 
As planets migrate outward, they may capture objects in orbital resonances. This process 
may yield some dynamical classes of trans-Neptunian objects \citep[e.g.,][]{mor08} and 
dense clumps of material in debris disks \citep[e.g.][]{wya03,wya05,mar07,cri2009}.

To conclude this section, Fig. \ref{fig:migratem} compares the variation of $\miso$ 
and $\mfast$ with semimajor axis for a plausible disk model.  We adopt a disk with
$\Sigma = \Sigma_0 a^{-1}$ and $\Sigma_0$ = 10~g~cm$^{-2}$ at 1 AU. Here, we assume 
a factor of 3 jump in the surface density of solid material at the snow line, $a_{snow}$ 
= 3~AU. We ignore the likely variation in the position of the snow line with time 
\citep{kk08}.  In this disk model, $\mfast$ ranges from 0.005 \mearth\ at 1~AU to 
$\sim$ 1 \mearth\ at 10~AU; 
$\miso$ grows from 0.07~\mearth\ at 1~AU to 10~\mearth\ at 10~AU.  Based on our simulations, 
low mass oligarchs with $m < \miso$ are too closely packed to migrate.  Prior to chaotic
growth, planetesimals can grow to reasonably massive oligarchs. With $m_{erode} > \miso$
at all $a$, disk erosion is also unimportant.  Once masses reach $\miso$, chaotic growth 
without migration leads to terrestrial mass planets at 1~AU \citep{ray05,kb06,ray09b} and 
Jupiter mass planets at 3--30~AU \citep{gol04,bk10}.

\section{Relationship to migration in gaseous disks}
\label{ssect:gaseousdisk}

To explore whether our results on migration are general, we now consider some analogies 
between migration in gaseous and planetesimal disks. As motivation, numerical simulations 
demonstrate that systems of many oligarchs are likely outcomes of runaway growth in a 
planetesimal disk. Although there are several elegant approaches to the migration of single 
planets in a gaseous disk \citep[see][and references therein]{pap07}, generalizing these
approaches to systems of 20--30 (or more) planets with masses comparable to or less than 
the isolation mass is challenging \citep[see][2008]{cre06}.  Here, we try to see whether 
we can apply results for planetesimal disks to gaseous disks.

In the limit of zero viscosity, equal mass gaseous and particle disks provide an identical
torque on an embedded planet \citep{gt80}.  In both types of disk, local variations in
density generate the torque. In a particle disk, scattering sets the density structure.
In a continuous medium, a balance between gravity, pressure, and viscous forces sets the 
density structure. 
As the viscosity of the medium increases, this structure damps out. In this heuristic picture,
planetesimals generate migration efficiently; a very viscous medium cannot generate migration.
However, the large mass of a gaseous disk gives it an overwhelming advantage over a planetesimal 
disk. For a solar metallicity system, the gaseous disk is roughly 100 times more massive than 
the disk of solids.

As predicted by \citet{gt80}, our numerical simulations produce coherent wakes from orbital
resonances close to embedded planets.  In planetesimal disks (Fig. \ref{fig:wake1}), an 
embedded planet scatters planetesimals out of the co-orbital zone into disk regions several 
$\rhill$ away from the planet. The region of horseshoe orbits is initially filled 
(as in Fig. \ref{fig:wake1}, left panel); continued scattering removes planetesimals from the 
co-orbital zone (as in Fig. \ref{fig:wake1}, right panel). In both panels, a bridge of enhanced 
planetesimal density connects the high density rings of planetesimals lying $\pm$ 4--5 
$\rhill$ away from the orbit of the planet. Because the 2:1 resonance lies outside the disk,
the strongest density enhancements lie at the 3:2, 4:3, and 5:4 resonances. 

To illustrate the time evolution of these structures, the online version of this paper
contains a movie of a planet migrating from 25 AU to 15 AU in $\sim 8 \times 10^4$ yr. 
Throughout the movie, the planet scatters planetesimals out of its orbit into various
resonances.
Fig. \ref{fig:wake2} shows a snapshot from the movie. At this point of the evolution,
the planet has migrated from 25 AU to $\sim$ 20 AU. At 25 AU, the original orbit of the planet
is nearly devoid of planetesimals. Just outside this orbit, the density of planetesimals 
is somewhat higher than the initial density. Between 20~AU and 25~AU, the planet has left 
behind a sea of stirred up planetesimals, with several density enhancements at orbital
resonances. At 20 AU, the planet has evacuated planetesimals downstream from its orbit.
Upstream, planetesimals remain in horseshoe orbits. 

The structures in planetesimal disks are similar to those produced in simulations of gaseous 
disks \citep[e.g.,][]{bate03,dang2003,kla06,paa06a}. In all of the simulations of planets
within gaseous disks, torques between the planet and the disk create local enhancements in 
the gas density at orbital resonances as well as the bridge of material from the planet to 
the bright rings.  For disks with similar surface density distributions and planets with 
similar masses, the derived range of the density enhancements are also similar 
\citep[see, for example][]{war97,dang2008}. Because gaseous disks have some pressure support, 
co-orbital gas lies a small distance, $\delta r_{co} \approx 0.002 - 0.004 a$, inside the 
orbit of a planet \citep[e.g.][]{tan02}.  When co-orbital gas lies inside the Hill sphere 
of the planet ($m \gtrsim $ 0.03 \mearth), we expect co-orbital gas and planetesimals with
similar surface density to exert similar torques on a nearby planet. Thus, gaseous and 
planetesimal disks should produce comparable migration rates for planets with $m \gtrsim m_{iso}$.

\subsection{Migration timescales}

To generalize our migration results to a gaseous disk, we consider the 
vertical scale height of the disk $h$ 
as a smoothing length, which sets the minimum size of density features in the disk. In Type I 
migration, this assumption limits the scale and the location of the density wakes that form 
through interaction with a relatively small planet.
The largest wakes lie at a distance $\dR \sim h \gg \rhill$. Thus, we can use results for 
small-angle scattering. For gaseous disks, the standard type I rate for an isothermal disk
from \citet{tan02} is
\begin{equation}
\dadtgasI =  
-(2.7 + 1.1 \powlaw) \frac{2 \pi a^2 \Sigma_g m}{M^2} ~ \frac{a^2}{h^2} ~ \frac{a}{T}\ ~ ,
\label{eq:typeIgas}
\end{equation}
where $\Sigma_g$ is the surface density of the gaseous disk at $a$.  Setting $\dR \sim h$ 
and assuming the planetesimals and the gas have the same surface density at the position 
of the planet, the ratio of the rate from eq.~(\ref{eq:dadtembedded}) to this rate is 
$\xi_I = 16(2 - \powlaw) / 3(2.7 + 1.1 \powlaw)$. Additional features in the \citet{tan02} 
derivation, including 3-D effects and corotation resonances, produce the different 
dependence on the surface-density power-law index $\powlaw$. For $\powlaw$ = 0.5--1.5, 
$\xi_I \approx$ 2--0.5; thus, the rates differ by a small numerical factor\footnote{Using 
the more recent type I rate from \citet{dang2010a} yields similar results.}.  Clearly, 
small-angle-scattering migration in a planetesimal disk and Type I migration in a gaseous 
disk share general properties.

Type II migration occurs when a relatively massive planet creates a gap in a gaseous 
disk and locks into the disk's viscous flow as a result of a build-up of material at 
the gap's edges.  Gravitational torques exerted on the planet by the disk produce 
inwards migration. Thus, the planet responds to the instantaneous density perturbations 
within the disk. These perturbations are strongest at the gap edges, which are several 
$\rhill$ away from the planet. The condition for gap opening is an elegant inequality
between $m$, $h$, and the disk viscosity parameter $\alpha$ \citep{cri2006}: 
\begin{equation}
\frac{3}{4} ~ \frac{h}{\rhill} + 50 \alpha ~ \frac{M}{m} ~ \left ( \frac{h}{a} \right )^2 \lesssim 1 ~ ,
\label{eq:gap}
\end{equation}
where $\alpha = \nu / h^2 \Omega$, $\nu$ is the disk viscosity, and $\Omega$ is the angular 
velocity \citep[e.g.,][]{pri81}.  This constraint has a simple physical interpretation. When 
the viscosity is small ($\alpha \rightarrow 0$), the first term dominates; planets with Hill 
radii comparable to the local disk scale height open a gap
\citep[see also][1993, and references therein]{war97,lin86}.  As the viscosity grows 
($\alpha \rightarrow 1$), the second term dominates; planets with tidal forces 
large enough to overcome viscous transport open a gap \citep[e.g.,][]{war97,bry99}.

From the condition for gap opening in the low viscosity limit and the \citet{tan02} type I 
migration rate, \citet{pap07} derive a simple estimate for the maximum type II rate.
For planets with $m \approx$ 30--1000 \mearth, their simple estimate agrees with 
rates derived from detailed numerical simulations \citep[see Fig. 3 of][]{pap07}.
Adopting $\dR \approx \rhill$ in the embedded migration timescale from 
eq.~(\ref{eq:tauemb2}) and assuming the same $\Sigma$, we derive a ratio of rates
$\xi_{II, \alpha = 0} = 32 (2 - \powlaw) / 3 (5.4 + 2.2 \powlaw) $ $\approx$ 2.5--0.6 for $\powlaw$ 
= 0.5--1.5.  Thus, both approaches yield the same scaling with $\rhill$ and magnitudes 
consistent to a factor of 2--3.  In this limit, irradiation from the central star sets
the scale height of the gas \citep[e.g.,][]{kh87,chiang1997}. Once this scale is set,
the mass of the planet establishes the region of the disk that interacts most strongly 
with the planet.  In the zero viscosity limit, planetesimals and gas respond to the 
gravity of the planet on the Hill scale, leading to similar timescales.

In the large viscosity limit, our analogy between planetesimal and gaseous disks breaks
down. Large viscosity enables a gaseous disk to transport mass inward and angular 
momentum outward. Viscous transport modifies the universal trajectory function $g(x)$. 
Identifying $g(x)$ for viscous transport is beyond the scope of this paper; however, 
we speculate that substituting proper expressions for $g(x)$ and the size of the gap 
in eq.~(\ref{eq:dadt}) would yield a migration rate reasonably close to published
type II rates, $da/dt \approx \alpha (h/a)^2 a \Omega$, where $\Omega$ is the angular 
velocity of the planet at semimajor axis $a$. Successfully applying our approach in the 
large $\alpha$ limit would link the theories of migration in gaseous and planetesimal
disks.

Type III migration is completely analogous to the fast migration mode
in planetesimal disks \citep[][see also eq. \ref{eq:mfast-min-gas}] {mas03}. 
Because it regulates how efficiently 
material is transported across a planet's orbital position, viscosity 
complicates precise comparisons between gaseous and particle disks.
However, viscosity generally makes migration in gaseous disks less 
efficient per unit disk mass than in planetesimal disks \citep{ida08}.
As a result, planets with masses much less than the mass of Saturn 
are `safe' from type III migration through the gaseous disk
\citep[e.g.,][2008b]{mas03, dang2005, pep08a}.

\subsection{Migration with multiple planets}

In a gaseous disk, there are three sources of torque on an embedded planet.  Torque from 
an inner spiral density wave and material in the corotation zone produces a net outward 
migration. Material in an outer spiral density wave causes a net inward migration.  As the 
planet migrates inward, viscous torques smooth out density perturbations behind the planet. 
Smoothing occurs on a local viscous timescale, which is comparable to the migration rate.  

In a multiple planet system, each planet produces a pair of spiral density waves.  Thus,
each planet feels a torque from the spiral density waves of all planets and the gas in 
its corotation zone. When planets are widely separated, distant spiral density waves 
contribute little to the torque. Widely spaced planets migrate freely. When 
planets are tightly packed, many spiral waves contribute to the torque. In linear theory
\citep[e.g.,][]{tan02}, multiple torques superpose and add to the migration. However, 
this approach does not address the response of the gaseous disk to the time-variable 
potential of a collection of closely packed planets. The gravitational potential of the 
planets varies on timescales shorter than the viscous timescale, which should wash out
spiral density waves and reduce migration rates.

Recent analyses show that the thermodynamics of the disk is an important factor in setting 
the direction and rate of type I migration \citep[e.g.,][]{paa08,paa10,paa11}. In these
non-linear calculations, migration of a single planet depends on the vertical temperature 
structure and the relative strength of torques from the corotation zone and the Lindbald
resonances. In a multiple planet system, each corotation zone generally lies within a few 
Hill radii of a single planet; thus, closely packed planets may not change the torque from 
the corotation zone. Because Lindbald resonances lie many Hill radii away from a planet, 
they are easily perturbed by an ensemble of closely packed planets which change the density 
and temperature structure on timescales shorter than the viscous timescale. Because the
disk responds relatively slowly to motions of the oligarchs, spiral density waves are
probably much weaker in a system with many oligarchs than in a system with a few oligarchs.
Weaker density waves produce smaller migration rates.  By analogy with our simulations of 
planetesimal disks, we propose that tightly packed planets do not migrate.

To place quantitative constraints on these limits, we compare the locations of 
the resonances that drive migration to the radial spacing of planets.  For type I 
migration, the gaseous disk produces the strongest torques at the inner and outer 
Lindblad resonances, which lie at orbital distances $\delta a_{LR} \approx$ 
$\pm 2 h / 3$ from the migrating planet \citep[e.g.,][]{pap07}. For two planets 
separated by $r \approx \pm 4 h / 3$, their Lindblad resonances overlap. 
This tight spacing may preclude the elegant sprial density waves necessary for 
type I migration. Planets separated by $r_{min} \approx 2 h$ have isolated 
Lindblad resonances and can migrate freely.  With 
$h \approx h_0$ (a/1 AU)$^{9/7}$ AU \citep{chiang1997}, this constraint becomes 
$r_{min} \approx$ 0.06 (a / 1 AU)$^{9/7}$~AU for $h_0$ = 0.03~AU.  To convert to 
Hill units, planets with $m \approx \mearth$ have $\rhill \approx 0.01 a$. Thus, 
our constraint is 
\begin{equation}
\label{eq:rmin}
r_{min} \approx 6 ~ \rhill ~ \left ( \frac{\mearth}{m} \right )^{1/3} \left ( \frac{a}{\rm 1~AU} \right )^{2/7} ~ .
\end{equation}

Numerical simulations do not yet address constraints on the ability of closely packed 
planets to undergo type I or type III migration.  \citet[][2008]{cre06} consider 
ensembles of Earth-mass or larger planets ($m \gtrsim \miso$) spaced by roughly 
5--7 $\rhill$.  In their simulations, type I migration is briefly interrupted by rapid, 
chaotic interactions among the planets.  Once the planets have merged or scattered, 
type I migration continues. Calculations for systems of lower mass planets with 
$m \lesssim \miso$ do not exist.  For now, we assume that ensembles of lower mass 
planets with typical separations smaller than $r_{min}$ do not migrate.

\subsection{Migration and planet formation}

To establish some constraints on type I migration through a gaseous disk containing an 
ensemble of growing planetesimals, we generalize our discussion of isolated oligarchs
from \S2.5.  
As planetesimals experience runaway, oligarchic, and chaotic growth, the gaseous disk evolves 
with time. In addition to viscous evolution, photoevaporation and gas giant planet formation 
remove mass from the disk \citep{ale09}. Observations of young stars suggest typical disk 
lifetimes of 1--3 Myr \citep{hai01,cur09,kenn09,mama09,wil11}.  
Thus, migration through the gaseous disk ceases after 1--3 Myr.

The lifetime of the gaseous disk places a rough lower limit on the masses of planets 
subject to type I migration. For the linear calculations of \citet{tan02}, the
timescale for type I migration is $\tau \propto m^{-1}$. Thus, lower mass planets 
migrate more slowly. Setting the timescale in eq~(9) of \citet{pap07} to 3 Myr, 
we can derive an expression for the masses of planets that are safe from 
type I migration:
\begin{eqnarray}\label{eq:msafe}
\msafe & \approx & 0.046 
  \left(\frac{h/h_0}{0.03}\right)^{2}
 \left(\frac{a}{\mbox{1 AU}}\right)^{\powlaw+1/14} 
\\
\ & \ & \times \ 
 \left(\frac{\Sigma_0}{1700\ \mbox{g~cm$^{-2}$}}\right)^{-1}
 \left(\frac{M}{1 \Msolar}\right)^{3/2}
\ \mearth ~ ,
\end{eqnarray}
where we assume a gaseous disk with $\Sigma = \Sigma_0 ~ a^{-\powlaw}$.
Planets with $m < \msafe$ migrate on timescales longer than the typical disk lifetime.
Fig.~\ref{fig:migratemg} compares the variation of $\msafe$ and $\miso$ with semimajor axis. 
For all $a$, $\miso \gg \msafe$. Low mass oligarchs have long migration timescales; 
isolated objects are not safe from type I migration.

Despite their lack of safety, many isolated oligarchs are packed too tightly together
to migrate. To draw this conclusion, we derive the masses of objects with $r_{min}$
= 7 $\rhill$, the typical separations of oligarchs at the onset of chaotic growth.
Defining $m_{min}$ as the mass where $r_{min}$ = 7 $\rhill$ 
\begin{equation}
\label{eq:mmin}
m_{min} \approx 0.63 \left ( \frac{a}{\rm 1~AU} \right )^{6/7} ~ \mearth ~ .
\end{equation}
In our picture, oligarchs with $m < m_{min}$ are packed too tightly to undergo type I
migration.  For $a \lesssim$ 5 AU, $m_{min} \gtrsim \miso$; isolated oligarchs have 
overlapping Lindblad resonances and are packed too closely to migrate.  At larger 
$a$, $m_{min} \lesssim \miso$; oligarchs do not have overlapping resonances and 
can migrate.

This result leads to an important conclusion for terrestrial planet formation. If
the overlapping Lindblad resonances of tightly packed oligarchs at 1 AU do not 
generate type~I migration, chaotic growth produces Earth mass or larger planets on 
timescales of $\sim$ 10~Myr.  In our simulations \citep[e.g.,][]{kb06}, it takes 
$\sim 3 \times 10^4$ yr ($\sim 10^5$ yr) to produce 5--10 ($\sim$ 15) oligarchs with 
$m \sim 0.002 - 0.004$~\mearth\ at 0.85--1.15~AU. These oligarchs are safe from type I 
migration through the gas (Fig. \ref{fig:migratemg}), but their low masses allow 
fast migration through the sea of leftover planetesimals.  However, growing oligarchs stir 
planetesimals to $\ehill \gae $ 5.  After migrating $\sim$ 0.02--0.03~AU, each oligarch 
encounters planetesimals stirred up by its inner neighbor. Relative to the standard fast 
migration rate, we estimate a factor of 10--100 reduction in the migration rate for 
each oligarch in our 2006 simulations.  On the (reduced) migration timescale of 
$\gtrsim 10^6$~yr, each oligarch in our simulations grows by more than an order of 
magnitude and begins to interact chaotically with other oligarchs. Once chaotic growth 
begins, oligarchs safely grow into terrestrial planets. 

As chaotic growth ends, several factors probably prevent terrestrial planets from 
migrating through the remnants of the gaseous disk or the sea of leftover planetesimals. 
In published simulations, leftover planetesimals have very large $e$ and $i$ 
\citep{ray05,kb06,obr06,ray09b}; thus, planets can sweep up or scatter the leftovers faster than 
they can migrate through them. For
typical disk lifetimes of 1--3 Myr, the reduced surface density lowers migration
rates through the gas by factors of 10 or more. Migration times are then longer 
than the disk lifetime, saving terrestrial planets from type I migration through 
the gas.

Formation outcomes for gas giant planets are less clear. In our picture, isolated
oligarchs at 5--10 AU will migrate little through a sea of leftover planetesimals.
As chaotic growth begins, these objects start to experience type I migration through 
the gas. The relative importance of chaotic growth and migration then depends on several 
factors.

\begin{enumerate}

\item The masses of leftover planetesimals.  Although oligarchs with gaseous atmospheres 
accrete small planetesimals rapidly \citep{ina2003,cha06a,bk10}, they cannot accrete large 
planetesimals on timescales shorter than the migration time \citep[e.g.,][2008]{cha06b}.  
Collisional grinding can reduce the sizes of large planetesimals, enabling rapid accretion
and the formation of 5--10 \mearth\ cores on very short timescales \citep{kb09}.  Thus,
rapid core formation depends on the evolution of the size distribution of planetesimals 
during oligarchic and chaotic growth.

\item The response of the disk to tightly packed oligarchs. When oligarchs are tightly 
packed, their Lindblad resonances overlap. By analogy with our calculations of migration
through a sea of planetesimals, we speculate that tightly packed oligarchs cannot migrate.
However, there is no analytic or numerical study of type I migration in gaseous disks with 
tightly packed oligarchs. If tightly packed oligarchs migrate at the `standard' type I rate,
then they migrate faster than they grow. If tightly packed oligarchs (and leftover 
planetesimals) reduce the type I migration rate, then they probably grow faster than they 
migrate.

\item The response of the disk to planets accreting gas. Once planets reach masses
of 1--10~\mearth, they begin to accrete material from the disk 
\citep{miz1980,ste1982,iko2000,raf2006,hor2010}.  At 5 AU, the nominal migration 
timescale for 10 \mearth\ planets is $\sim 5 \times 10^4$ yr, shorter than the 
accretion timescale of $\gtrsim 10^5$ yr \citep{pol96,bod00,kor02,pap07,bk10}.  However, 
this planet may not migrate so quickly. When the size of the corotation zone is comparable
to the disk scale-height, the disk may not be able to launch coherent density waves for 
Type I migration. If corotation torques are important, migration may stall until the planet 
reaches larger masses, forms a gap in the disk, and begins Type II migration \cite{mas06}.

\end{enumerate}

\section{Discussion}

Fig. \ref{fig:migratesum} summarizes the main conclusions of our analysis. When
planets grow in a planetesimal disk (left panel), interactions between closely 
packed oligarchs ($m < \miso$) or between chaotic oligarchs ($m > \miso$) limit 
migration through a sea of planetesimals. Thus, the building blocks of terrestrial
planets and ice or gas giant planets are safe from this form of migration.  

In a gaseous disk (right panel), we speculate that low mass planets ($m < m_{min}$) are packed 
too closely to undergo type I migration. If this constraint is correct, the building blocks 
of terrestrial planets rarely undergo type I migration. Once they are fully formed, 
terrestrial planets can migrate through the disk. However, the reduced surface 
density of the disk then limits migration to small radial distances. 

Even with these constraints, type I migration is still an issue for the building 
blocks of gas giant planets. At 5--10 AU, gas giant planet formation depends on
the relative importance of migration and chaotic growth. If chaotic growth 
dominates, the cores of gas giants can form before they migrate. If migration
dominates, planets must accrete enough material to begin to accrete gas before 
they migrate into the central star.

Improving these conclusions requires a better understanding of the transition
from oligarchic growth to chaotic growth.  During the early stages of oligarchic 
growth, oligarchs are closely packed within a fairly uniform sea of stirred up 
planetesimals embedded in a fairly uniform gaseous disk. As oligarchs grow, they 
become more and more isolated. As they become isolated, oligarchs push the excited 
planetesimals out of their orbits \citep[e.g.,][]{raf2001}. This evolution creates 
two types of density perturbations within the disk. 

\begin{itemize}

\item At the onset of chaotic growth, closely-packed oligarchs contain roughly 
50\% of the mass of solids.  These oligarchs create point-like density enhancements
in the surface density distribution of the solids.  

\item Planetesimals contain the other half of the solid material in the disk. 
Planetesimals tend to concentrate in rings between the orbits of the oligarchs.

\end{itemize}
Thus, the surface density distribution of the solids is fairly rippled, with
planetesimals concentrated in the peaks of the ripples and oligarchs orbiting 
within the troughs of the ripples.

Current theory addresses the response of the gaseous disk to isolated oligarchs.
For a standard viscous disk, analytic results and numerical simulations yield 
reasonably robust solutions to the structure of a gaseous disk with an ensemble 
of widely spaced oligarchs \citep[e.g.,][]{pap07,cre08,lub10}.  Despite many 
remaining uncertainties in treating the (thermo)dynamics of the gas, the eccentricity 
of the oligarchs, magnetic fields, turbulence, and other phenomena, interactions 
between isolated oligarchs and the disk clearly lead to migration.

Although the planetesimal theory predicts ensembles of closely packed oligarchs,
migration theory does not address the structure of the gaseous disk at the onset
of chaotic growth.  Closely packed oligarchs clearly cannot migrate through a 
planetesimal disk (Fig.~\ref{fig:multi}). We speculate that overlapping Lindblad 
resonances prevent migration through a gaseous disk. New analytic and numerical 
approaches are required to test this idea.

Migration theory also does not include the response of the disk to the ensemble 
of leftover planetesimals. Analytic solutions suggest oligarchs create gaps in 
the surface density distribution of planetesimals \citep{raf2001}. Many numerical 
simulations show that growing oligarchs push away and scatter leftover planetesimals
\citep[e.g.,][and references therein]{mal93,kok98,mor08,kir09}. With $\sim$ 50\% of 
the solid mass at the onset of chaotic growth, structure in the spatial distribution
of planetesimals probably leads to density waves within the gas. Density waves from
individual planetesimals probably have negligible impact on oligarchs or planetesimals.
However, density waves from the ensemble of planetesimals can interact with oligarchs 
orbiting several $\rhill$ away.  It is not clear whether this interaction impacts 
migration significantly; however, including the behavior of planetesimals is necessary 
for a complete theory of migration through a gaseous disk.

Addressing the response of the disk to closely packed oligarchs and to leftover
planetesimals will improve our understanding of planet formation. Despite our 
good working knowledge of the growth of oligarchs from planetesimals
\citep[e.g.,][2010]{wet93,kb08}, the formation of planetesimals 
\citep[e.g.,][]{you2010}, the transition from oligarchy to chaos 
\citep[e.g.,][]{gol04,kb06}, and the long-term evolution of fully-formed 
planets within a gaseous disk \citep[e.g.,][]{ida05,pap07,lub10} are less 
robust aspects of the theory.  Complete numerical simulations of migration 
with a gaseous disk, closely-packed oligarchs, and a sea of leftover planetesimals 
are beyond the capabilities of current computers. Smaller simulations of disks 
with rings of planetesimals and a few oligarchs are possible and would begin to 
address how planetesimals might change migration rates through the disk.

\section{Summary}

We have used analytic results and numerical simulations to explore aspects of 
migration in protostellar disks.

\begin{figure}[htb]
\centerline{\includegraphics[width=5.5in]{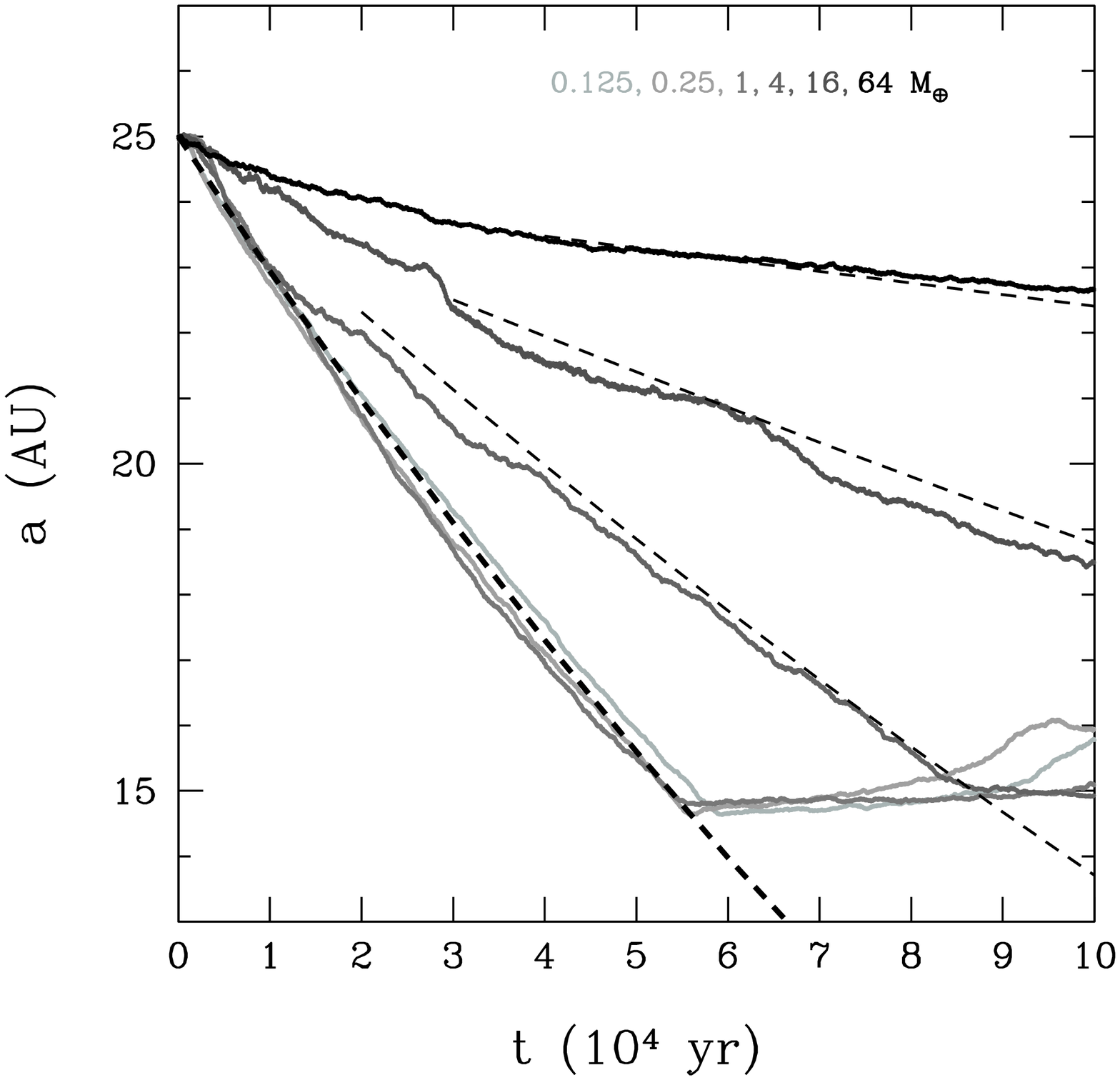}}
\caption{%
  Migration of planets with different masses through a planetesimal disk.  
  The disk is a sea of equal-mass particles, with 
  $\Sigma = 1.2 (a/{\rm 25~AU})^{-1}$ g cm$^{-2}$, extending from 14.5~AU 
  to 35.5~AU \citep[e.g.][]{kir09}. Planetesimals have masses of 1/600$^{\rm th}$ 
  the mass of the planet and initial r.m.s. eccentricity of $1\ \ehill$.  The 
  three lowest mass planets with $m < \mfast \approx 3\ \mearth$ undergo fast 
  migration (heavy dashed curve; eq.~[\ref{eq:dadtfast}]), until they ``bounce'' 
  off the inner edge of the disk. More massive planets migrate more slowly 
  (light dashed curves), at a rate that scales with $\mfast/m$).
\label{fig:migratemass}}
\end{figure}

\begin{figure}[htb]
\centerline{\includegraphics[width=5.5in]{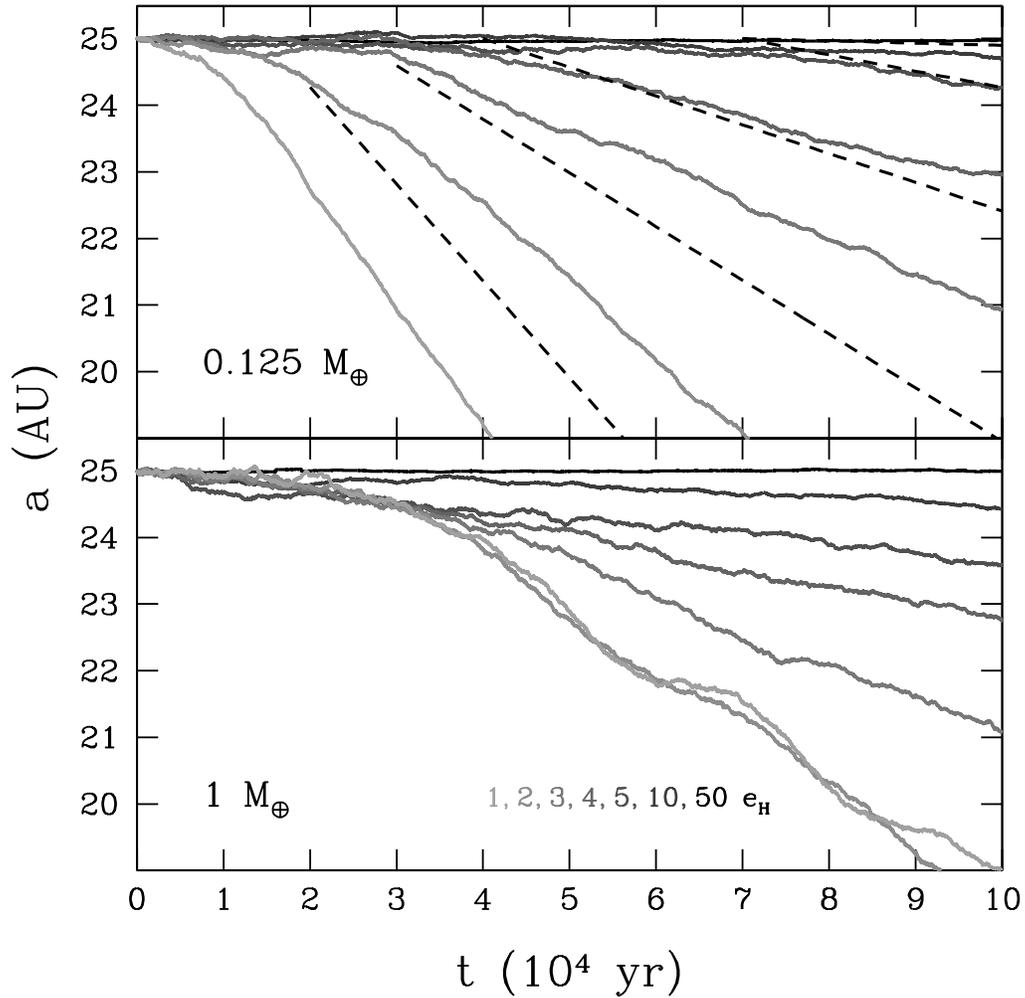}}
\caption{%
  Planetary migration in disks with $\Sigma$ as in Fig.~\ref{fig:migratemass}
  and various values for the initial r.m.s.\ eccentricity.  
  The lower left corner of each panel indicates the mass of the planet.
  In each panel, the initial $\ehill$ varies from 1 to 50 as indicated in the
  legend of the lower panel (lighter shades correspond to smaller initial 
  eccentricity).  The dashed curves in the upper panel show predicted rates 
  from eq.~(\ref{eq:migrateecc}).
\label{fig:migrateecc}}
\end{figure}

\begin{figure}[htb]
\centerline{\includegraphics[width=5.5in]{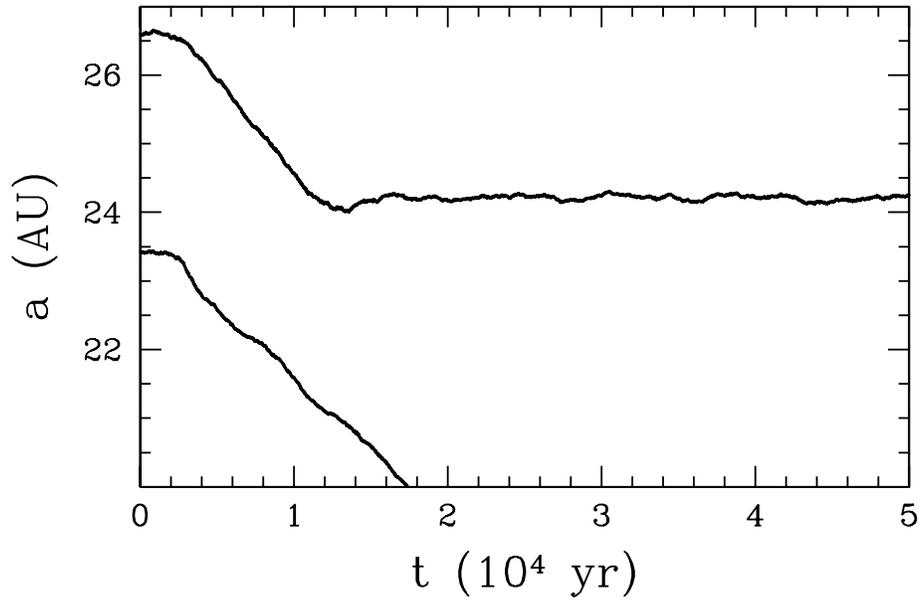}}
\caption{%
  Disrupted migration of a planet as a result of stirring by its neighbor.  The 
  planetesimal disk has initial conditions as in Fig.~\ref{fig:migratemass}.
  Two 0.5~\mearth\ planets have initial separation of 16~$\rhill$.  Each begins 
  fast migration.  The inward motion of the outer planet stops when it encounters 
  the wake of excited planetesimals left behind by the inner planet.
\label{fig:pairfast}}
\end{figure}

\begin{figure}[htb]
\centerline{\includegraphics[width=5.5in]{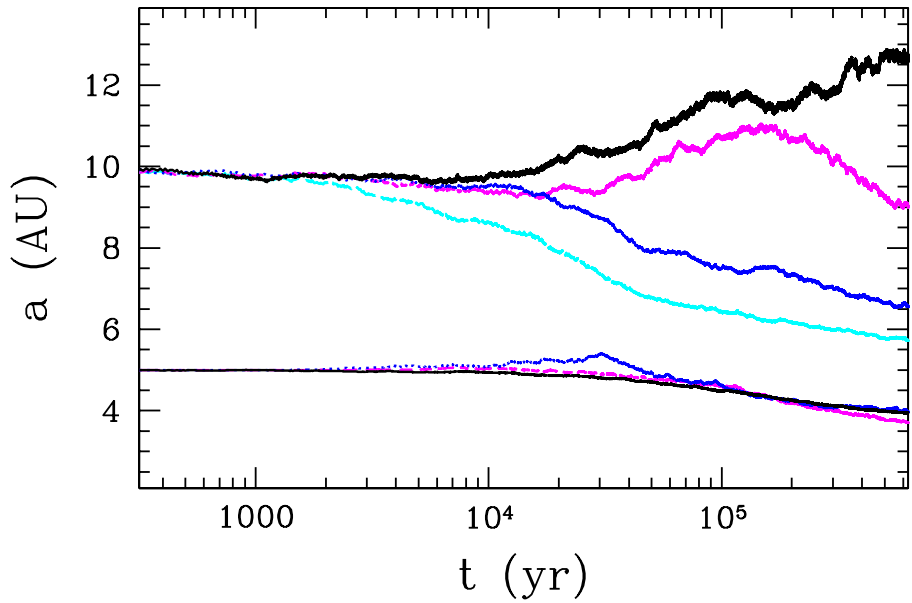}}
\caption{%
  Migration of a Saturn-mass planet in a planetesimal disk with initial
  conditions as in Fig.~9 of \citet{lev07}. The disk extends from 6~AU 
  to 20~AU and has a mass equal to the combined mass of Jupiter and Saturn.
  The Saturn mass planet begins at 10~AU. The inner planet starts at 5 AU 
  and has the mass of Jupiter (black), Saturn (magenta) and 30 \mearth\ (blue). 
  The cyan curve shows the outer planet migrating through the disk in the 
  absence of any other planet.
\label{fig:migrateJS}}
\end{figure}

\begin{figure}[htb]
\centerline{\includegraphics[width=5.5in]{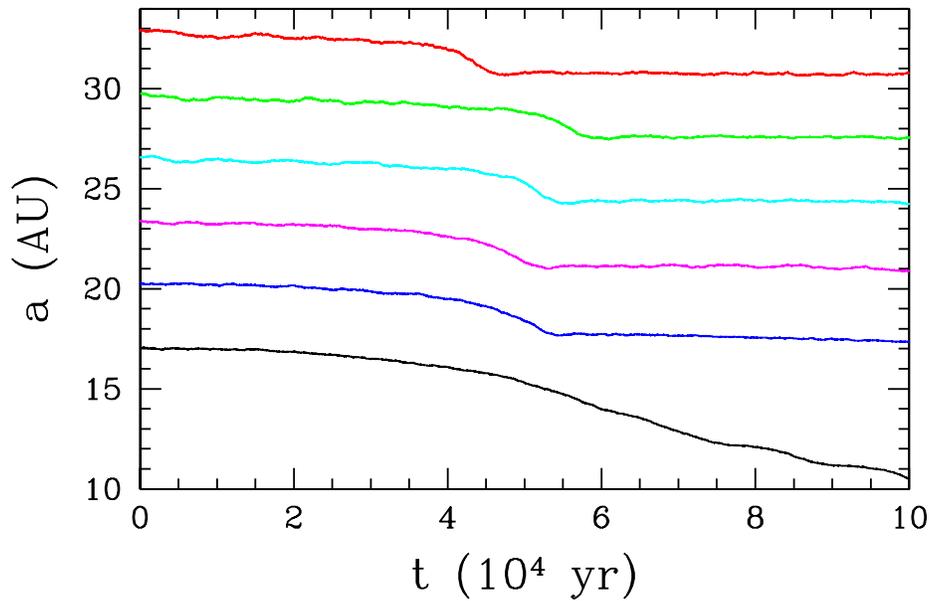}}
\caption{%
  Migration of multiple planets in a planetesimal disk with initial conditions
  as in Fig.~\ref{fig:migratemass} except that the inner edge of the disk is at
  5~AU.  Six 0.5~\mearth\ planets separated by 16~$\rhill$ migrate inward until 
  they encounter the wakes of their inner neighbor. Once the innermost planet
  reaches $\sim$ 10~AU, its mass exceeds $\mfast$. Because $\mfast \propto a^{3/2}$,
  its migration rate then slows dramatically.
\label{fig:multi}}
\end{figure}

\begin{figure}[htb]
\centerline{\includegraphics[width=5.5in]{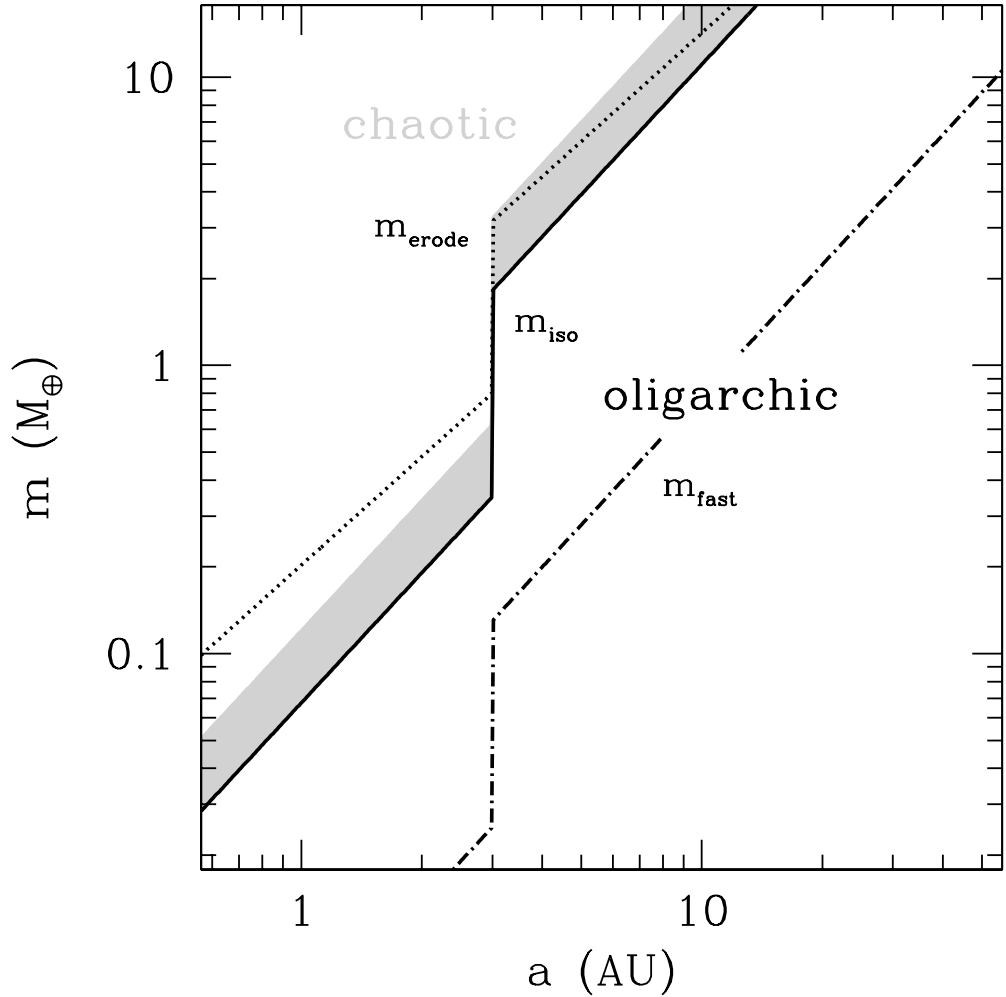}}
\caption{%
Growth and migration modes in a planetesimal disk. The heavy solid curve and
the shaded region indicate the variation of the isolation mass ($\miso$) with 
semimajor axis; planets with $m < \miso$ ($m > \miso$) undergo oligarchic 
(chaotic) growth. The dot-dashed curve indicates the variation of $\mfast$ with
$a$; planets with $m < \mfast$ ($m > \mfast$) undergo fast (slow) migration. 
Until planets reach $\miso$, they are tightly packed and unable to migrate large 
distances through the disk.  Once they have $m > \miso$, they are free to migrate 
in the slow mode. As planets grow larger than $\miso$, their migration may be 
slowed by disk erosion, as indicated by the dotted line ($m_{erode}$).
\label{fig:migratem}}
\end{figure}

\begin{figure}[htb]
\centerline{\includegraphics[width=6.0in]{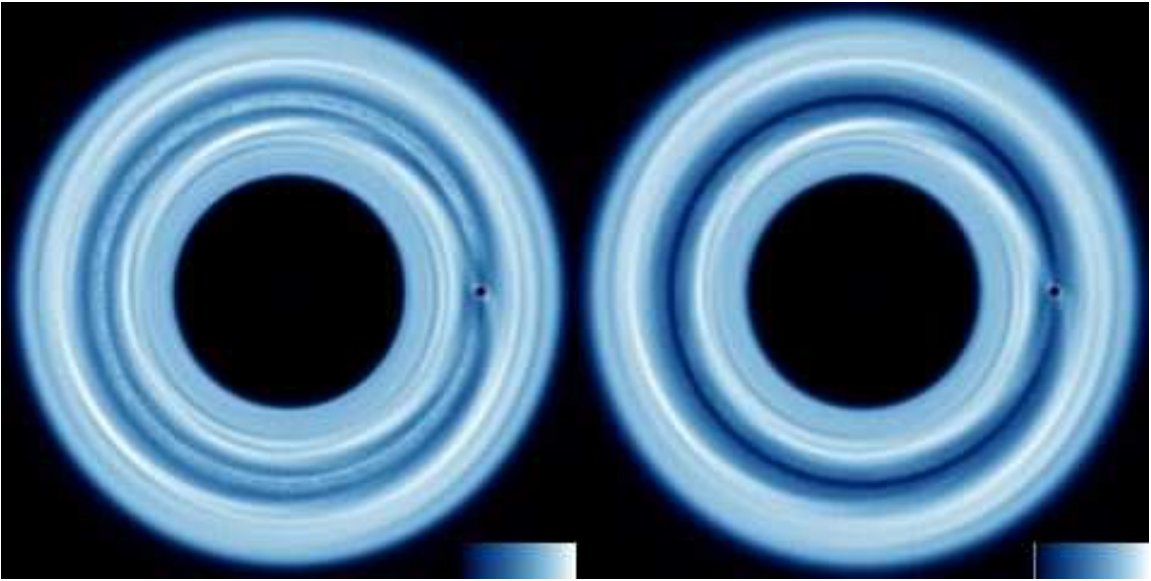}}
\caption{%
Density wakes in planetesimal disks with an embedded planet.  Both images
are in a frame rotating with the planet, which has a mass of 16~\mearth\ 
and a semimajor axis of 25 AU.  In the left panel, the corotation zone contains 
planetesimals; in the right panel, the corotation zone is empty.  From the inner 
edge of the disk at $\sim$ 15~AU
to the outer edge at $\sim$ 35~AU, the images show the local planetesimal density 
-- averaged over 1~kyr -- relative to the initial surface density,
$\Sigma(a) = 30~{\rm g~cm^{-2}}~(a/{\rm 1~AU})$. In the lower right corner of 
each image, the scale shows the linear map of density to color. The full range
of the color map is a factor of two in the local mean density.  
\label{fig:wake1}}
\end{figure}

\begin{figure}[htb]
\centerline{\includegraphics[width=6.0in]{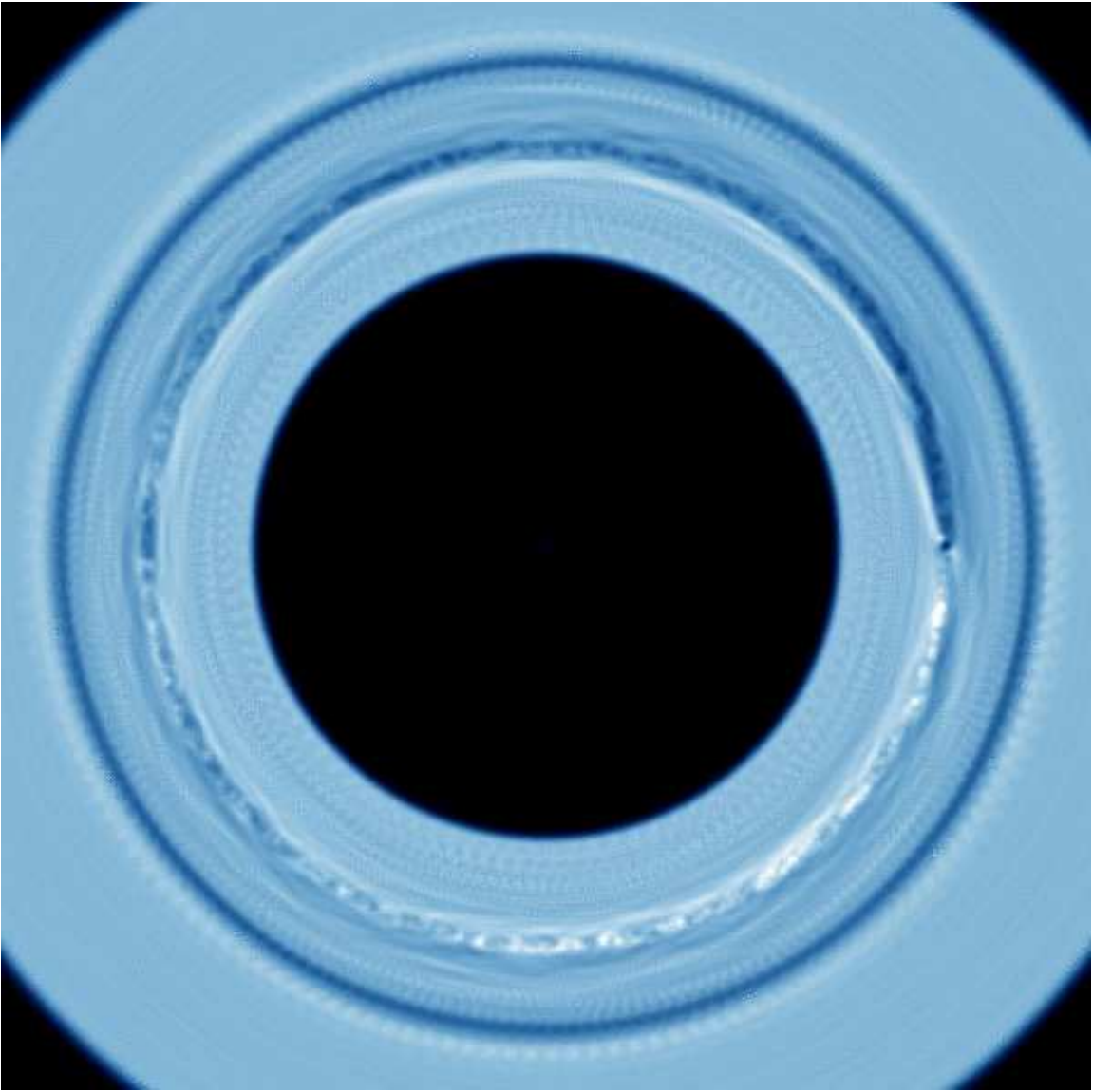}}
\caption{%
As in Fig.~\ref{fig:wake1} for a planet experiencing fast migration.  This image is a 
snapshot from a simulation of a 1~\mearth\ planet, available in the on-line version of 
the Journal. The simulation shows density structures after the planet has moved several 
AU inward from its initial orbital distance at 25 AU. When the planet is in fast migration 
mode, the upstream corotation zone is filled; the downstream region is relatively empty.
\label{fig:wake2}}
\end{figure}

\begin{figure}[htb]
\centerline{\includegraphics[width=6.0in]{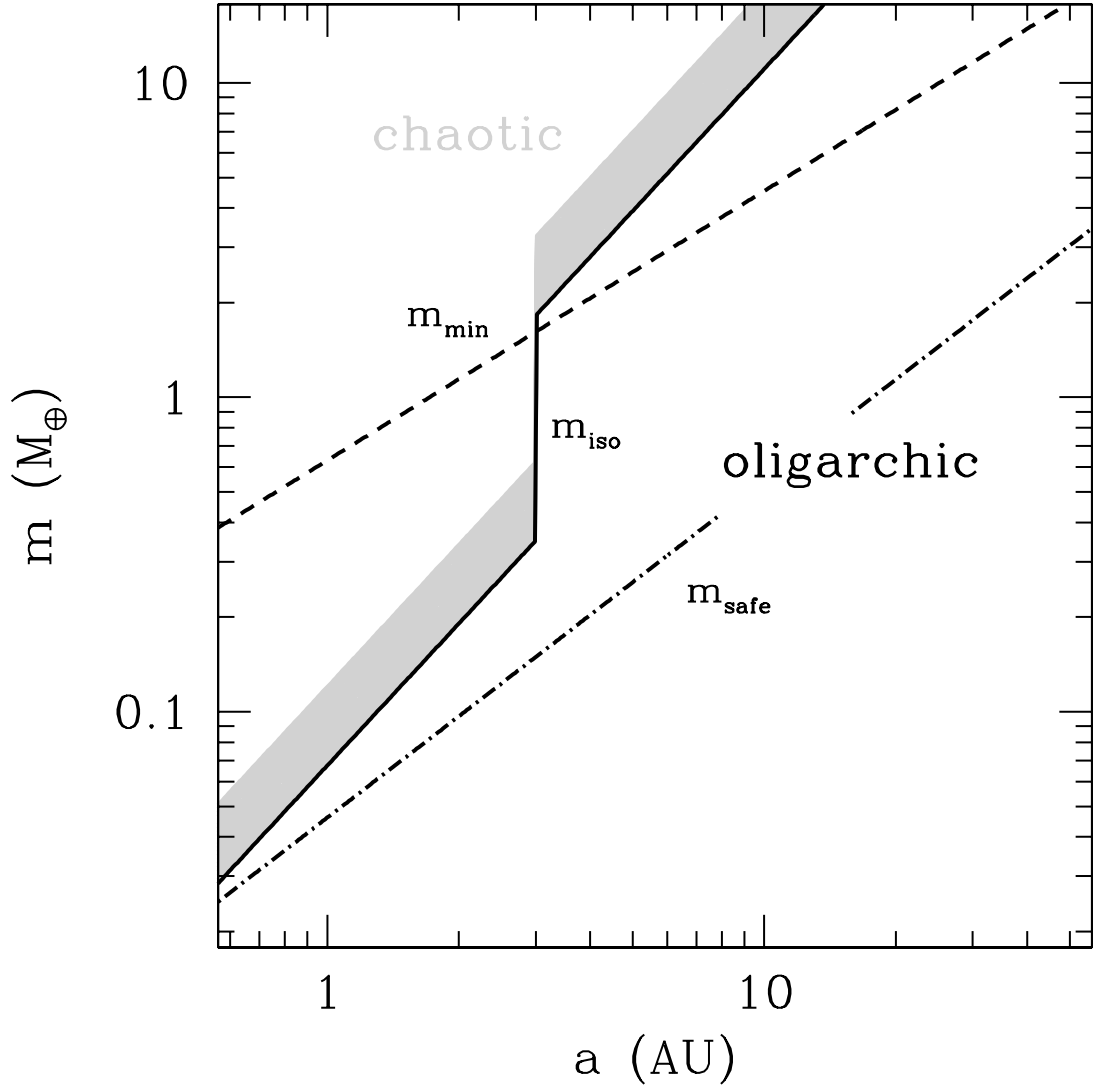}}
\caption{%
As in Fig.~\ref{fig:migratem} for a gaseous disk. Curves for isolation mass are 
also from Fig.~\ref{fig:migratem}. Planets with $m < m_{safe}$ (dot-dashed line)
migrate on timescales longer than the lifetime of the gaseous disk.  Before they 
migrate significantly, the gas disperses. Planets with $m < m_{min}$ (dashed
line) are packed too closely to migrate through the gaseous disk. Terrestrial 
planets likely undergo chaotic growth before they are able to migrate. The cores
of gas giant planets start to migrate as they begin chaotic growth.
\label{fig:migratemg}}
\end{figure}

\begin{itemize}

\item We derive `universal' rates for isolated planets migrating rapidly, 
eq. (\ref{eq:dadtfast}), or slowly, eq. (\ref{eq:dadtembedded}), through a disk 
of planetesimals.  When the mass of the planet is much much smaller than the
mass of the central star, these rates agree with comprehensive numerical 
simulations and with rates derived from previous studies 
\citep[e.g.,][]{ida00,lev07,kir09,bk10}. We derive an upper limit $\mfast$ 
(eq. (\ref{eq:mfast})) on the mass of a rapidly migrating planet. In a disk with 
surface density $\Sigma$ = 30 g cm$^{-2}$ at $a$ = 1 AU, $\mfast \approx$ 
0.025~\mearth; for $\Sigma \propto a^{-1}$, $\mfast \propto a^{-3}$.  When
$m > \mfast$, fast migration rates are inversely proportional to the mass of 
the planet (Fig.~\ref{fig:migratemass}). This result is new.

\item Tests of planets migrating through a disk of stirred up planetesimals verify 
that rates scale with the eccentricity of background planetesimals in Hill units,
$\ehill^{-3}$ \citep[Fig.~\ref{fig:migrateecc}; see also][]{ida00,kir09}.
 
\item The strong scaling with $\ehill$ suggests that planets cannot migrate 
through the wakes of stirred up planetesimals left behind by another migrating
planet. Several tests confirm this hypothesis (Figs. \ref{fig:pairfast}--\ref{fig:multi}). 
Thus, closely-packed oligarchs do not migrate. This result is also new.

\item When a newly-formed planet migrates or is scattered into a region where 
planetesimals have small $\ehill$, this isolated planet can migrate through a 
large part of the disk \citep[see also][]{mal93,lev07}.

\end{itemize}

We use some simple arguments to generalize these results to migration through a
gaseous disk.

\begin{itemize}

\item Adopting the disk scale height $h$ as the scale for density perturbations
in the disk, we show that rates for type I, type II (in the zero viscosity limit), 
and type III migration through gaseous disks are similar in magnitude and scaling 
to rates through planetesimal disks.

\item If closely-packed oligarchs migrate as poorly through gaseous disks as they
migrate through planetesimal disks, we derive limits on the masses of oligarchs 
that undergo type I migration through disks with surface density 
$\Sigma = \Sigma_0 a^{-1}$. 

\end{itemize}

\begin{figure}[htb]
\centerline{\includegraphics[width=5.5in]{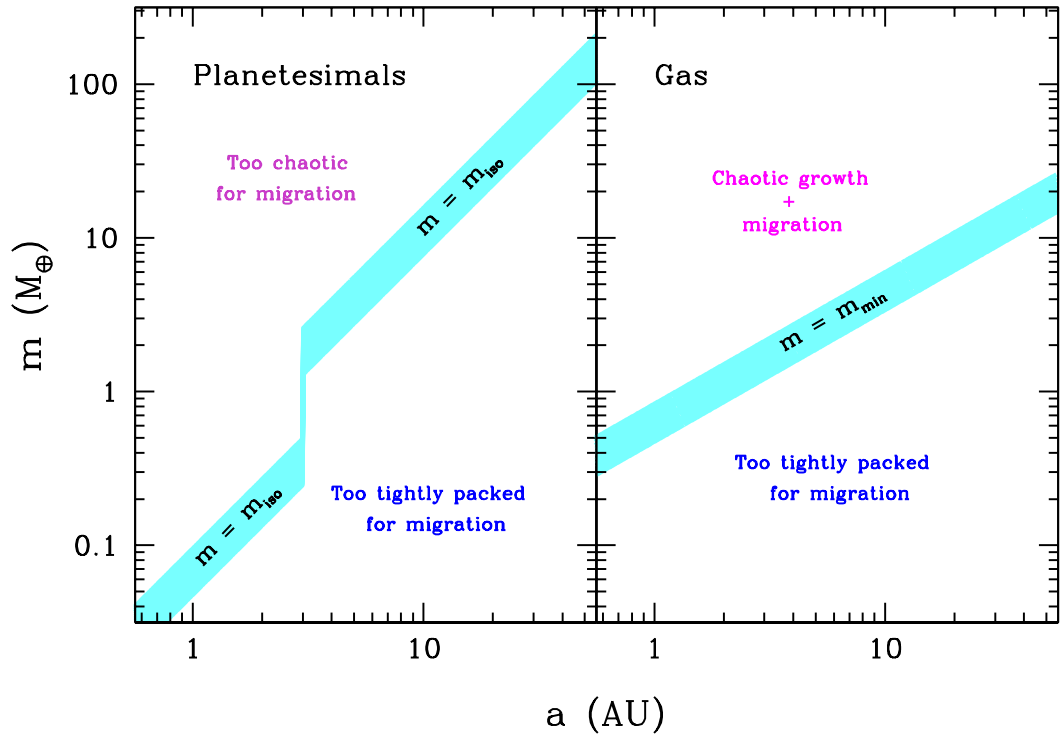}}
\caption{%
Migration in gaseous and planetesimal disks.
In a planetesimal disk (left panel), planets with $m < \miso$ are packed too
closely to migrate. When $m > \miso$, chaotic growth dominates migration.  In 
a gaseous disk (right panel), planets are spaced too closely to migrate when 
$m < m_{min}$. Once $m > m_{min}$, planets grow chaotically as they migrate.
The relative importance of chaotic growth and migration probably depends on
the response of the disk to smaller oligarchs and leftover planetesimals.
\label{fig:migratesum}}
\end{figure}

Combining these results into a single diagram (Fig.~\ref{fig:migratesum}), we 
conclude that type I migration is an important issue during the formation of
gas giant planets. The building blocks of these planets are probably safe until 
they reach the isolation mass ($\miso$; eq. (\ref{eq:miso})). Once their masses 
exceed $\miso$, the migration rate depends on how the gas responds to the mass 
distribution of smaller oligarchs and leftover planetesimals. Addressing this 
issue requires new analyses.

For terrestrial planets, we conclude that type I migration is unimportant.  Throughout 
oligarchic and chaotic growth, the building blocks of rocky planets are packed too 
closely to migrate. Once these planets are fully-formed, the surface density of the 
gas is probably too low to support type I migration. Thus, our analysis suggests
that standard calculations of terrestrial planet formation without migration yield
robust estimates of the formation timescale and orbital properties of terrestrial 
planets.

\acknowledgements

Advice and comments from M. Duncan, M. Geller, D. Kirsh, S. Tremaine, A. Youdin,
and an anonymous referee greatly improved our presentation.  Portions of this 
project were supported by {\it NASA's } {\it Astrophysics Theory Program,} and 
the {\it Origin of Solar Systems Program} through grant NNX10AF35G.


\end{document}